\documentclass{aa}  
\usepackage{graphicx}
\usepackage{natbib}
\usepackage{booktabs}
\bibpunct{(}{)}{;}{a}{}{,}
\usepackage{txfonts}
\usepackage[colorlinks=true,citecolor=blue,linkcolor=blue]{hyperref}
\usepackage{comment}
\usepackage{caption}
\usepackage{multicol}
\usepackage{subcaption}
\usepackage{mhchem}
\usepackage{academicons}
\usepackage{xcolor}
\usepackage{svg}
\usepackage{orcidlink}
\usepackage{newtxtext}
\usepackage[varvw]{newtxmath}
\newcommand{\rt}[1]{{#1}}

\newcommand{\micron}{\mu \mathrm{m}}
\newcommand{\amin}{a_{\mathrm{min}}}
\newcommand{\amax}{a_{\mathrm{max}}}
\newcommand{\Rin}{R_{\mathrm{in}}}
\newcommand{\Rout}{R_{\mathrm{out}}}

\newcommand{\mcmax}{\texttt{MCMax3D}}
\newcommand{\prodimo}{\texttt{ProDiMo}}
\newcommand{\Rtap}{R_{\mathrm{tap}}}
\newcommand{\Mdust}{M_{\mathrm{dust}}^\mathrm{cav}}
\newcommand{\Mgas}{M_{\mathrm{gas}}^\mathrm{cav}}
\newcommand{\dd}{\mathrm{d}}

\newcommand{\Mjup}{M_{\mathrm{Jup}}}

\newcommand{\Mearth}{M_{\oplus}}

\newcommand{\Msun}{M_{\odot}}

\newcommand{\kabs}{\kappa_{\nu}^{\mathrm{abs}}}
\newcommand{\kabsave}{\langle \kappa_{\nu}^{\mathrm{abs}} \rangle}
\newcommand{\kb}{k_{\mathrm{B}}}

\newcommand{\Tb}{T_\mathrm{b}}
\newcommand{\Mp}{M_\mathrm{p}}
\newcommand{\Ms}{M_*}

\newcommand{\co}{\ce{^{13}CO}}  % 13^CO 
\newcommand{\coo}{\ce{C^{18}O}} % C^18O
\newcommand{\cooo}{\ce{^{12}CO}} % 12^CO 
\newcommand{\CO}[3]{\ce{^{#1}CO $J=#2-#3$}}  % x^CO J=i-j
\newcommand{\COO}[3]{\ce{C^{#1}O $J=#2-#3$}} % C^xO J=i-j
\newcommand{\Sigmax}[1]{\Sigma_\mathrm{#1}} 

\newcommand{\hp}{h_\mathrm{p}} 

\newcommand{\gofish}{\texttt{GoFish}}

\begin{document}

   %\title{Constraining the gas and dust distribution in the PDS 70 disk using thermo-chemical models}
   \title{Constraining the gas distribution in the PDS 70 disk as a method to assess the effect of planet-disk interactions}
   %\subtitle{I. Overviewing the $\kappa$-mechanism}

   \author{B. Portilla-Revelo \inst{1} \orcidlink{0000-0002-6278-9006},
          I. Kamp \inst{1} \orcidlink{0000-0001-7455-5349}, 
          S. Facchini \inst{2,6}\orcidlink{0000-0003-4689-2684}, 
          E. F. van Dishoeck\inst{3,7} \orcidlink{0000-0001-7591-1907},
          C. Law \inst{5}\orcidlink{0000-0003-1413-1776},
          Ch. Rab \inst{4,7} \orcidlink{0000-0003-1817-6576},
          J. Bae \inst{8} \orcidlink{0000-0001-7258-770X}, 
          M. Benisty \inst{9} \orcidlink{0000-0002-7695-7605}, 
          K. \"Oberg \inst{5} \orcidlink{0000-0001-8798-1347},
          and R. Teague \inst{10} \orcidlink{0000-0003-1534-5186}
          }
    \authorrunning{B. Portilla-Revelo et al.} 
    \titlerunning{Constraining the gas distribution in the PDS 70 disk}
    
   \institute{Kapteyn Astronomical Institute, University of Groningen,
              PO Box 800, 9700 AV Groningen, The Netherlands \\
              \email{bportilla@astro.rug.nl}
        \and
            Dipartimento di Fisica, Università degli Studi di Milano, Via Celoria 16, 20133 Milano, Italy
        \and
            Leiden Observatory, Leiden University, PO Box 9513, 2300 RA Leiden, The Netherlands
        \and
            University Observatory, Faculty of Physics, Ludwig-Maximilians-Universität München, Scheinerstr. 1, 81679 Munich, Germany   
        \and       
            Center for Astrophysics | Harvard \& Smithsonian, 60 Garden St., Cambridge, MA 02138, USA
        \and
            European Southern Observatory, Karl-Schwarzschild-Str. 2, D-85748 Garching, Germany
        \and
            Max-Planck-Institut f\"ur Extraterrestrische Physik, Geissenbachstrasse 1, 85748 Garching, Germany
        \and
            Department of Astronomy, University of Florida, Gainesville, FL 32611, USA
        \and
            Univ. Grenoble Alpes, CNRS, IPAG, 38000 Grenoble, France
        \and
            Department of Earth, Atmospheric, and Planetary Sciences, Massachusetts Institute of Technology, Cambridge, MA 02139, USA
             }

   \date{Accepted XXX}
   %\date{}
    
% \abstract{}{}{}{}{} 
% 5 {} token are mandatory

  % aims heading (mandatory)
   
  \abstract
  % context heading (optional)
  % {} leave it empty if necessary  
   {Embedded planets are potentially the cause of substructures like gaps and cavities observed in the continuum images of several protoplanetary disks. Likewise, the gas distribution is expected to change in the presence of one or several planets and the effect can be detected with current observational facilities. Thus, the properties of the substructures observed in the continuum and in line emission encode information about the \rt{presence of planets in the system and how they interact with the natal disk}. The pre-transitional disk around the star PDS 70 is the first case of two young planets imaged within a dust depleted gap that was likely carved by themselves.}
  % aims heading (mandatory)
   {We aim to determine the spatial distribution of the gas and dust components in the PDS 70 disk. \rt{The axisymmetric substructures observed in the resulting profiles are interpreted in the context of planet-disk interactions.}}
  % methods heading (mandatory)
   {We develop a thermo-chemical forward model for an axisymmetric disk to explain a subset of the Atacama Large Millimeter/Submillimeter Array (ALMA) band 6 observations of three CO isotopologues \rt{plus the continuum towards PDS 70. The model accounts for the continuum radiative transfer, steady-state chemistry, and gas thermal balance in a self-consistent way and produces synthetic observables via ray-tracing.}}
  % results heading (mandatory)
   {We demonstrate that the combination of a homogeneous dust size distribution across the disk and relatively low values of the viscosity ($\alpha \lesssim 5\times 10^{-3}$) can explain the band 6 continuum observations. For the gas phase, analysis of the synthetic observables points to a gas density peak value of ${\sim} 0.1 \ \mathrm{g}\ \mathrm{cm}^{-2}$ located at $75$ au and a minimum of ${\sim} 10^{-3} \ \mathrm{g}\ \mathrm{cm}^{-2}$ at $20$ au. The location of the minimum matches the semi-major axis of the innermost planet PDS 70 b. Combining the gas and dust distributions, the model results in a variable gas-to-dust ratio profile throughout the disk that spans two orders of magnitude within the first $130$ au and shows a step gradient towards the outer disk, which is consistent with the presence of a pressure maxima driven by planet-disk interactions. Particularly, the mean gas-to-dust ratio within the dust gap between $16-41$ au is found to be ${\sim} 630$. \rt{We find a gas density drop factor of $\mathbf{{\sim} 19}$ at the location of the planet PDS 70 c with respect to the peak gas density at $75$ au. Combining this value with literature results on the hydrodynamics of planet-disk interactions, we find this gas gap depth to be consistent with independent planet mass estimates from infrared observations.} \rt{Our findings point towards gas stirring processes taking place in the common gap due to the gravitational perturbation of both planets.}} 
   {\rt{The distribution of gas \rt{and dust} in the PDS 70 disk can be constrained by forward modelling the spatially resolved observations from high resolution and sensitivity instruments like ALMA. This information is a key piece in the qualitative and quantitative interpretation of the observable signatures of planet-disk interactions.}}


   \maketitle
%
%-------------------------------------------------------------------
 
\section{Introduction}
\label{sec:intro}
The accepted explanation for the origin of giant planets points to a formation process that takes place in the midplane of disks made of gas and dust around young stellar objects. In consequence, a planet's properties are expected to be regulated, at least to a first order, by the amount and the properties of the material the disk is made of. Therefore, it is logical to think of a two-sided relationship where the properties of the planet and the disk are determined by each other. Observing young planets in the embedded phase is thus a crucial step to understand how their properties link to those of the disk.   

The observability of planets embedded in the natal protoplanetary disk was elusive until recently due to the observational limitations to image planets close to the host star. Typical working angles achieved with Hubble Space Telescope are close to ${\sim} 1$ arcsec, whereas ground based telescopes lack the sensitivity needed to spot faint structures. Several attempts to observe embedded protoplanets have covered a broad region of the spectrum, from short (visual and near-infrared) to long (submillimeter) wavelengths \citep{Benisty2022}. In the short wavelength regime, particularly at visual wavelengths, accretion signatures can be observed. When the star is actively accreting, inflow of gas from the outer disk has to make its way to the inner disk, and along its journey, part of it can be accreted by the planets resulting in accretion signatures like $\mathrm{H}\alpha$ emission (e.g. \citealt{Natta2004,Szulagyi2020}). Searches for those signatures have been carried out with scarce success within the solar neighbourhood \citep{Zurlo2020}. At near infrared wavelengths, the high continuum optical depth prevents planets from being directly imaged; hence gaps and cavities are suitable places to search for protoplanets due to their relatively low extinction \citep{Paardekooper2004,Espaillat2014}. Recent advances on differential imaging techniques not only have allowed the identification of disk substructures but have also led to the detection of young planets since they emit thermal emission at those wavelengths (see \citealt{Benisty2022} for a recent review). At submillimeter wavelengths, spatially resolved gas kinematics is another way to detect embedded planets (see \citealt{Pinte2022} for a review). The presence of a planet can induce deviations from a Keplerian profile in the isovelocity curves of the gas component. The extent of the deviation is proportional to the mass of the planet causing the distortion and it can be detected with interferometric facilities like ALMA. Based on this idea, two planets with a few times the mass of Jupiter have been proposed to explain the perturbations in the velocity field and substructures observed in HD 163296 \citep{Pinte2018,Teague2018} and in HD 97048  \citep{Pinte2019}. In both cases, the location of the planets would lie beyond 100 au; this reflects a limitation of the gas kinematic method due to ALMA not being able to probe the gas kinematics inside ${\sim} 10$ au.   

Recent advances in high spatial resolution and high contrast optical and near-infrared imaging of disks allowed the discovery of two protoplanets in the cavity of a pre-transitional disk: PDS 70 b and c \citep{Keppler2018,Haffert2019}. Infrared spectroscopy of the planets alongside analysis of their astrometric data constrain their masses down to a few times that of Jupiter and suggest a 2:1 resonant orbital configuration \citep{Muller2018,Wang2021}. Observations in $\mathrm{H}\alpha$ showed that the planets are accreting circumstellar material at a relatively low rate of ${\sim} 10^{-2} - 10^{-1}\, \Mjup\ \mathrm{Myr}^{-1}$ \citep{Haffert2019,Hashimoto2020} indicative of a post-runaway growth scenario. This also suggests that there is still gas flowing through the dust depleted cavity. In a recent ALMA campaign, \cite{Facchini2021} identified emission from 12 molecular species in the PDS 70 circumstellar disk, including three CO isotopologues, $\mathrm{H}_2\mathrm{CO}$, and small hydrocarbons. They found that most of the emission comes from the outer disk while the region within the orbit of the innermost planet displays only $^{12}\mathrm{CO}$ and $\mathrm{HCO}^+$ emission. 

There have been several attempts to determine the mass of the two protoplanets in the PDS 70 disk. This is particularly challenging for planet c because of its closeness to the outer dust ring and the presence of a circumplanetary disk, both acting as sources of contamination. For planet b, \cite{Hashimoto2020} inferred a value of $12\ \Mjup$ analysing the properties of the $\mathrm{H}\alpha$ line. \cite{Keppler2018,Muller2018,Mesa2019,Wang2020} found values between $1$ and $17\ \Mjup$ using near-infrared spectroscopy. Finally, \cite{Bae2019,Wang2021} placed an upper limit of $10\ \Mjup$ via orbital stability considerations. \rt{A possible way to ameliorate these uncertainties is using detailed thermo-chemical models able to translate multiple observational constraints into physical properties (such as density distributions) that encode information about the planets' mass.}

Hydrodynamical simulations show that the presence of one or several massive protoplanets can partially deplete the disk from primordial material through accretion and/or transfer of angular momentum in the form of spiral density waves \citep{Paardekooper2022}. For the single-planet case,  \cite{Kanagawa2015} found a scaling relation between the level of gas depletion in the cavity and the mass of the planet. This relation also depends on the disk's properties, particularly on the scale height and on the disk's viscosity. \cite{Duffell2015} performed a similar study including up to three planets in the simulations. They found that when planets are massive enough, their spheres of influence are consequently larger, favouring gap merging. Once this regime has been reached, the mutual gravitational perturbation among the planets stirs the gas inside the common gap, making it shallower. This implies that in order to get the same level of gas depletion created by a single planet of a given mass, a system with multiple planets would require them to be more massive. With these simulations being computationally demanding, \cite{Duffell2015} explored only a limited subset of the parameter space. Particularly, that work is restricted to equal mass planets in a 2:1 orbital commensurability within a disk characterised by a Shakura-Sunyaev $\alpha$ parameter of $10^{-2}$. In conclusion, simulations suggest that the level of gas depletion in a cavity carved out by one or several planets encodes information about their mass.

\rt{All of the above suggests that a robust constraint on the spatial distribution of gas and dust can inform us about the intricate connection among planets and disk. In this work, we retrieve the gas and dust distribution in the PDS 70 disk and analyse the inferred substructures in the context of planet-disk interactions.} We use a subset of the ALMA gas observations from \cite{Facchini2021} imaged at high spatial resolution, gas and dust radiative transfer forward modelling, and semi-analytic results from hydrodynamical simulations. In Section \ref{sec:methods} we describe the observations, the tools, and the methods we use to develop the thermo-chemical model. In Section \ref{sec:results} we compare the synthetic observables to the observations, we present the radial dependence of the gas-to-dust ratio, and \rt{quantify the amount of gas in the cavity}. \rt{In Section \ref{sec:discusion} we apply the derived radial profiles to compute the level of gas depletion in the gap. We use this information to estimate a characteristic value for the mass of the planets and to qualitatively assess the stirring of gas produced by them. In this section, we also discuss the main caveats of our approach.} Our conclusions are listed in Section \ref{sec:conclusions}.

\section{Methods}
\label{sec:methods}
\subsection{Observational data}
\label{sec:observations}
 We work with ALMA band 6 observations taken as part of the program \#2019.1.01619.S (PI S. Facchini). The instrument configuration was targeted to detect spatially-resolved emission from molecular lines at 0.1 arcsec resolution covering the frequency range 217.194-233.955 GHz. Sixteen transitions were detected from twelve different molecular species including isotopologues. In particular, we use the imaged set for \COO{18}{2}{1}, \CO{13}{2}{1}, and \CO{12}{2}{1} presented in Law et al. (in prep) with a beam FWHM of ${\sim} 0.15$ arcsec ($17$ au) and an rms noise level of ${\sim} 1.0$ $\mathrm{mJy}\ \mathrm{beam}^{-1}$ estimated on the mean value of the spectrum.  %\red{Note that our analysis is restricted to ALMA band 6 frequencies because those are expected to be less optically thick and therefore a better density tracer. The present work does not contain a detailed analysis of band 7 frequencies from which extra information on the disk temperature can be gained. Nevertheless we show that the model reproduces the spatially resolved observation in band 7 of $\CO{12}{3}{2}$ within a reasonable approximation.} 

A full description of the observations and data reduction process is presented in \cite{Facchini2021} and Law et al. (in prep). 

\subsection{Radiative transfer simulations: from the continuum to gas-phase models}
\label{sect:rt_model}
The two-dimensional model for the continuum presented in \cite{Portilla-Revelo2022} is taken as a starting point to develop a thermochemical model for the source. That model reproduces the ALMA observation at $855 \, \micron$ and the VLT/SPHERE observation in polarised-scattered light at $1.25\, \micron$. 

The model for the dust emission was developed in the three-dimensional version of the continuum radiative transfer code \mcmax \ \citep{Min2009}\footnote{\mcmax\ is a continuum radiative transfer code based on the Monte Carlo method. It solves the temperature structure of a distribution of dust particles assuming a full three-dimensional geometry implemented on a spherical grid where multiple sources of illumination can be included.}. To translate it into the thermochemical code \prodimo \ \citep{Woitke2009,Kamp2010,Thi2011,Woitke2016}\footnote{ProDiMo is a radiation thermochemical code that solves for the continuum radiative transfer (using a discrete ordinate method instead of a Monte Carlo approach), the gas heating-cooling balance, and the chemical abundances. The code assumes a two-dimensional geometry implemented on a cylindrical grid. In this work, we use \prodimo \ version \texttt{ebd47ddf}. The code can be found in \href{https://prodimo.iwf.oeaw.ac.at/}{https://prodimo.iwf.oeaw.ac.at/}}, we keep the properties of the dust disk constant as given in Table \ref{tab:model_properties}. The protoplanetary disk is simulated as a two-zone object with a boundary at $41$ au that delimits the transition from the gap's outer edge to the outer disk in the continuum. Both zones are populated with dust grains made of carbon and silicates in percentages by volume of $19.2\%$ and $60.8\%$, respectively, and with a fraction of vacuum accounting for $20\%$ of the grain volume. The 2D dust mass density distribution was mapped from the spherical grid of \mcmax \ onto the cylindrical grid of \prodimo \ to subsequently derive the column density per grain size by integrating over the vertical coordinate. The cylindrical grid has 120 radial and 100 vertical points logarithmically spaced. Figure \ref{fig:Sigma_dust} shows the total dust surface density as well as some values for selected grain sizes. The dust column density is used to define the limits of the dust gap which we assume to extend from $16$ au to $41$ au. This profile also sets an upper limit of $10^{-5} \, \mathrm{g}\, \mathrm{cm}^{-2}$ to the dust column density inside the gap indicating a drop by a factor of $2500$ with respect to the outer disk. The dust model is limited by the ALMA continuum sensitivity at $855 \micron$ which explains why it is radially extended only up to $130$ au. 

In \prodimo, the Shakura-Sunyaev $\alpha$ parameter \citep{Shakura1973} mimics dust settling according to \cite{Dubrulle1995}; in essence, the $\alpha$ value sets a different scale height to each bin size in the dust distribution. Each dust scale height is proportional to the gas pressure scale height and the proportionality constant varies inversely with the grain size. \cite{Portilla-Revelo2022} used a value of $\alpha \ {\sim} 10^{-2}$ across the disk which is equal to the value fixed in the simulations by \cite{Duffell2015} and consistent with the upper limit explored by \cite{Kanagawa2015}. However, recent numerical studies of viscous evolution of disks \citep{Delage2022} as well as empirical evidence \citep{Rosotti2023} support values of $\alpha$ lower than a few times $10^{-3}$. In order to have a compromise between these two scenarios, we adopted in this work a constant value of $\alpha=5\times 10^{-3}$ in both zones. This is within the range of the existing constraints  presented in \cite{Rosotti2023} for other disks. A lower turbulence enhances dust settling which in turn lowers the submillimeter flux, especially from the outer disk where most of the dust mass resides. Therefore, to compensate for a lower submillimeter flux, we homogenised the dust size distribution across both zones to be in the range $0.001 \ \micron$ to $3\ \mathrm{mm}$. This modification decreases the dust opacity of the inner disk compared to that in \cite{Portilla-Revelo2022} allowing more photons to reach the outer disk which increases the heating and boosts the submillimeter flux. A final fine-tune to the surface density profile was required to reproduce the ALMA continuum image at $855\ \micron$ and the polarised scattered image at $1.25 \ \micron$. This change produces a disk dust mass of $3.5\times 10^{-5}\ \Msun$, that is $22\%$ less than the mass estimated in \cite{Portilla-Revelo2022}. Both profiles are compared in Fig.\ref{fig:sigma_dust_compara_PR2022}. A comparison of the midplane dust temperature for both models is shown in Fig. \ref{fig:Td_midplane}.  

For the gas component, we use a parameterised structure for the vertical distribution. We assume a Gaussian vertical profile: 

\begin{equation}
\label{eq:gas_density}
\rho(r,z) \propto \exp{\bigg(-\frac{z^2}{2H_\mathrm{g}(r)^2}\bigg)}\, ,
\end{equation}

\noindent where the parametrised gas scale height is: 

\begin{equation}
\label{eq:gas_hscale}
H_\mathrm{g}(r)=H_0 \ \Bigg(\frac{r}{r_0}\Bigg)^{\beta}.
\end{equation}

\noindent Here, $H_0$ is the reference scale height at $r_0$ and $\beta$ is the flaring exponent.   

For completeness, the full set of model parameters and their values are summarised in Table \ref{tab:model_properties}. With the parameterised structure of the gas disk and the dust column density, \prodimo \ calculates the continuum radiative transfer and iterates over the heating-cooling balance and the chemistry in the disk. For the chemistry we use a reduced chemical network with $100$ species and $1286$ reactions \citep{Kamp2017}. In our model, CO isotopologue chemistry is not included and consequently the effects of isotope-selective photodissociation of CO are not simulated. The ratios for $^{12}\mathrm{CO}/^{13}\mathrm{CO}$  and $^{12}\mathrm{CO}/\mathrm{C}^{18}\mathrm{O}$ are taken from \cite{Henkel1994}. In the line radiative transfer step, we simulate the following transitions: \COO{18}{2}{1}, \CO{13}{2}{1}, and \CO{12}{2}{1}. 

% One column table
\begin{table*}
\caption{Parameters of the thermochemical model}
\label{tab:model_properties}
\centering
\begin{tabular}{lll}
\toprule \toprule
\multicolumn{3}{c}{Stellar parameters}\\
\midrule
Distance & $113.47$ pc\\
Mass & $0.76\, \mathrm{M}_\odot$\\
Bolometric luminosity ($L_\mathrm{bol}$) & $0.35\, \mathrm{L}_\odot$\\
Effective temperature & $3972\, \mathrm{K}$\\
$L_\mathrm{UV}/L_\mathrm{bol}$ & $4\times 10^{-4}$ \tablefootmark{(a)} \\
\midrule
Property & Zone 1\tablefootmark{(b)} & Zone 2 \\
\midrule
\multicolumn{3}{c}{Physical properties of dust grains}\\
\midrule
Minimum particle size: $\amin \, (\micron)$ & 0.001 & 0.001\\
Maximum particle size: $\amax \, (\micron)$ & 3000 & 3000\\
Particle size power index: $p$ & 3.9 & 3.9\\
Porosity (\%) & 20 & 20\\
Amorphous carbon by volume (\%) & 19.2 & 19.2\\
Turbulent settling parameter: $\alpha$ & $5\times 10^{-3}$ & $5\times 10^{-3}$\\
\midrule
\multicolumn{3}{c}{Geometrical properties}\\
\midrule
Inner radius: $\Rin$ (au) & 0.04 & 41\\
Outer radius: $\Rout$ (au) & 41 & 130\\
Flaring exponent: $\beta$ & 1.14 & 1.14\\
Reference radius: $r_0$ (au) & 100 & 100\\
Scale height at $r_0$: $H_0$ (au) & 10.61 & 10.61\\
\bottomrule
\end{tabular}
\tablefoot{\tablefoottext{a}{The value of the UV luminosity is estimated using the results from \cite{Yang2012} relating the far-ultraviolet luminosity to the accretion luminosity of the star (see Table 6 in that paper). The accretion luminosity is evaluated using a characteristic value of the mass accretion rate of $1.4 \times 10^{-10} \, \Msun \ \mathrm{yr}^{-1}$ (e.g. \citealt{Haffert2019,Thanathibodee2020}) alongside Eq. (8) in \cite{Gullbring1998}.} \tablefoottext{b}{Zone 1 is the computational domain that hosts the inner disk plus the dust-depleted gap while zone 2 contains the outer disk.}}
\end{table*}

 \begin{figure*}[h!]
\centering
\begin{subfigure}[b]{0.45\textwidth}
\includegraphics[width=\linewidth]{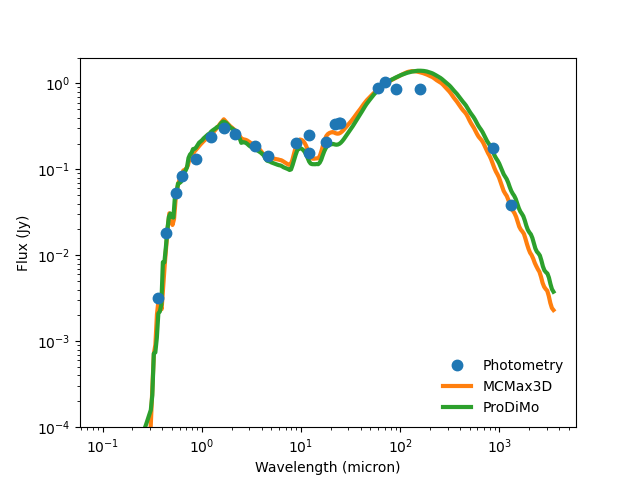}
%\caption{\small Predicted spectral energy distribution (blue) and photometry towards PDS 70 (orange).}
\label{fig:sed}
\end{subfigure}
%\hfill 
\begin{subfigure}[b]{0.45\textwidth}
\includegraphics[width=\linewidth]{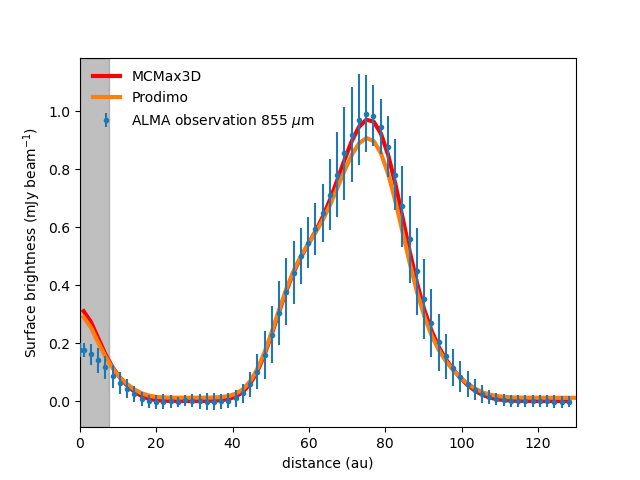}
%\caption{\small Predicted spectral energy distribution (blue) and photometry towards PDS 70 (orange).}
\label{fig:sed}
\end{subfigure}
\caption{\small (Left): predicted spectral energy distribution (orange and green curves) and photomety towards PDS 70 (blue dots). The photometric data were taken from \cite{Gregorio-Hetem1992}, \cite{Cutri2003}, \cite{Hashimoto2012}, and \cite{Keppler2019}. The value at $855 \ \micron$ \citep{Keppler2019} results from integrating the brightness distribution  over an elliptical aperture of projected radius $130$ au. (Right): azimuthally averaged radial profiles from the ALMA continuum observation at $855 \, \micron$ and from the modelled images ray-traced at the same wavelength. The shaded area indicates the resolution element of the continuum observation  of $0.067$ arcsec.}
\label{fig:sed}
\end{figure*}

\subsection{Post-processing the model and comparing to the observation}
\label{sect:post-processing}
The simulated CO data cubes were post-processed as follows. First, we use CASA version 6.4 \citep{McMullin2007} to convolve the cube with a Gaussian beam. The size and orientation of the resolution element passed to the \texttt{convolve2d} task to match the observation can be found in Table \ref{tab:obs_properties} for each transition. Then, we use the \texttt{imcontsub} task to subtract the continuum signal from each channel in the convolved cube; one-third of the total number of channels were assumed to be line-free and were used to fit a zero-order polynomial to estimate the continuum level. Finally, we use the task \texttt{immoments} to collapse the cube and create the moment-zero map for each transition; in this step we use all the continuum-subtracted channels in the cube without any sigma clip on the pixel values along the spectral axis. 

The azimuthally-averaged radial profile of the moment-zero map is obtained using the \gofish\ package \citep{Teague2019}. In this step, we adopted for the disk an inclination of $i=51.7^\circ$ with respect to the plane of sky and a position angle of $160.4^\circ$ measured east-of-north \citep{Keppler2019}. The moment zero map was binned radially with a spacing of $\theta_\mathrm{maj}/4$, where $\theta_\mathrm{maj}$ is the major axis of the resolution element of the observation. 

The resulting radial profiles of the synthetic moment zero maps are overlaid to the observations and the goodness of the model is determined by visual inspection.  

\begin{table*}
\caption{Properties of the observational data used in this work as presented in Law et al. in prep}
\label{tab:obs_properties}
\centering
\begin{tabular}{lll}
\toprule \toprule
Transition & Rest frequency & Beam: $\theta_\mathrm{maj} \times \theta_\mathrm{min}$ (PA)\\
\midrule
\CO{12}{2}{1} & $230.537939$ & $0.13'' \times 0.10'' \ (-83.75^\circ)$\\
\CO{13}{2}{1} & $220.398682$ & $0.15'' \times 0.12'' \ (-79.95^\circ)$\\
\COO{18}{2}{1} & $219.560296$ & $0.16'' \times 0.13'' \ (-82.40^\circ)$\\
\bottomrule
\end{tabular}
%}
%\tablefoot{\tablefoottext bla}
\end{table*}

\begin{comment}
\begin{table*}
\caption{Properties of the observational data set}
\label{tab:obs_properties}
\centering
\begin{tabular}{llll}
\toprule \toprule
Transition & Rest frequency & Beam: $\theta_\mathrm{maj} \times \theta_\mathrm{min}$ (PA) & Reference\\
\midrule
\CO{12}{2}{1} & $230.537939$ & $0.13'' \times 0.10'' \ (-83.75^\circ)$ & Law et al. in prep.\\
\CO{13}{2}{1} & $220.398682$ & $0.15'' \times 0.12'' \ (-79.95^\circ)$ & Law et al. in prep\\
\COO{18}{2}{1} & $219.560296$ & $0.16'' \times 0.13'' \ (-82.40^\circ)$ & Law et al. in prep\\
\bottomrule
\end{tabular}
%}
\tablefoot{\tablefoottext{a}{bla}}
\end{table*}
\end{comment}

\begin{comment}
\begin{equation}
\label{eq:surface_density}
\Sigma_{\mathrm{dust}}(r)=\Sigma_0 \bigg(\frac{\Rtap}{r}\bigg)^\epsilon\exp\Bigg[-\bigg(\frac{r}{\Rtap}\bigg)^{2-\gamma}\Bigg], 
\end{equation}

\noindent where $\Rtap$ is the tapering-off radius indicating the distance where the transition from a linear to an exponential decay of the surface density takes place. We use $\epsilon=\gamma=1$. The constant $\Sigma_0$ is determined by the total dust mass through the relation:
\end{comment}

\subsection{Modelling optically thin lines}
\label{subsection:fit_thin}
This section introduces the method we use to reproduce observed integrated intensities from optically thin emitters. It is based on an iterative procedure where the initial guess is provided by a simple semi-analytic model that connects the optically thin emission from an adequate tracer with the total gas column density. 

First we need to determine a suitable optically thin tracer. From the solution of the radiative transfer equation with no background sources, the ratio of the integrated intensities for \co\ and \coo\ would be equal to the ratio of the respective optical depths if both isotopologues were optically thin. Let us assume that the optical depth ratio is equal to the respective isotopologue ratio in the local ISM of $7.3 \pm 0.7$ \citep{Wilson1994}. Then, we would expect the integrated intensities to scale roughly as $I_{\co}\ \Delta v / I_{\coo}\ \Delta v \approx 7$. However, the observed profiles suggest that there are regions in the disk where this ratio is close to 3, too low to be consistent with the uncertainties in the isotopologue ratio measurement. This suggests that \co\ emission is moderately optically thick and accordingly, we assume \coo\ as a suitable proxy for optically thin emission.  

The intensity emitted by a plane-parallel slab of optically thin material can be related to the population of emitters in the upper excited state $N_i$ via:

\begin{equation}
\label{eq:N_up}
N_i=\frac{8 \pi\ \nu_{ij}^2\ \kb}{A_{ij}\ h\ c^3}\int \Tb \, \dd v,
\end{equation}

\noindent where $\nu_{ij}$ and $A_{ij}$ are the frequency and the spontaneous emission coefficient of the transition $i \rightarrow j$, $\kb$ and $h$ are the Boltzmann and Planck constants, $c$ is the speed of light, and $\Tb$ is the brightness temperature integrated in intervals of velocity $\dd v$. Therefore for optically thin \COO{18}{2}{1}, the observed brightness temperature can be linked to the abundance of \coo \ molecules in the upper state. Additionally, if local thermodynamic equilibrium (LTE) holds, the level population of the excited state can be written explicitly in terms of the total abundance of \coo \ via the partition function. Equation (\ref{eq:N_up}) can be finally expressed in terms of the \coo \ surface density $\Sigmax{C18O}$ as   

\begin{equation}
\begin{split}
\label{eq:Tb_vs_Sigma}
\int \Tb \, \dd v =& \ \frac{1.70\times 10^9}{\cos i} \ \Bigg(\frac{\Sigmax{C18O}}{\mathrm{g}\ \mathrm{cm}^{-2}} \Bigg) \
\Bigg(\frac{\mathrm{K}}{T_\mathrm{exc}(r)} \Bigg) \ \times\\  
&\exp\Bigg(\frac{-15.9\ \mathrm{K}}{T_\mathrm{exc}(r)}\Bigg) \ \mathrm{K\ km\ s}^{-1},
\end{split}
\end{equation}

\begin{comment}
\begin{equation}
\label{eq:Tb_vs_Sigma}
\int \Tb \, \dd v = \frac{3.8\times 10^7}{\cos i} \ \Bigg(\frac{\Sigmax{C18O}}{\mathrm{g}\ \mathrm{cm}^{-2}} \Bigg) \ \mathrm{K\ km\ s}^{-1},
\end{equation}
\end{comment}

\noindent where $\cos i$ is a correction factor that accounts for the inclination of the source with respect to the plane of sky. For simplicity, the position-dependent excitation temperature $T_\mathrm{exc}(r)$ is assumed to be equal to the dust temperature at the corresponding midplane location (see Fig. \ref{fig:Td_midplane}). 

Combining the semi-analytic model given by Eq. (\ref{eq:Tb_vs_Sigma}) with the observed azimuthally-averaged profile of the moment zero map for \COO{18}{2}{1}, we obtain an initial approximation to the column density distribution of \coo \ molecules. To estimate the total gas column density from the \coo\ distribution, we have to make a series of assumptions about the isotopologue abundance with respect to that of $\mathrm{H}_2$. As we are only interested in getting an initial approximation of the gas distribution, we adopt the local interstellar medium value for the ratio $\cooo/\coo \approx 560$ \citep{Wilson1994} alongside a canonical $\cooo/\mathrm{H}_2 \approx 10^{-4}$ \citep{Dickman1978}. The resulting semi-analytic profile for the total gas distribution is shown in Fig. \ref{fig:Sigma_gas_compara} and we note that it already points to the presence of a gas depleted gap. 

We use this initial gas distribution to run a first full thermochemical \prodimo \ model (i.e. solving for the continuum radiative transfer, heating-cooling balance, and chemistry) which is post-processed following the steps described in Sect. \ref{sect:post-processing}. Although the synthetic observables retrieved from the first run do not explain the observations satisfactorily, the semi-analytic approach is still useful because we do not require to perform an expensive exploration of the parameter space that defines the gas column density. Nonetheless, the semi-analytic approach ignores complexities like any cumulative effect on the heating-cooling balance and on the chemistry that could arise from the column of material built up in the radial direction and that is consistently modelled in a full thermochemical simulation.

Subsequently, we modify the gas distribution using the following iterative procedure. After running the first thermochemical model using the semi-analytic initial gas distribution, the radial profile of the synthetic moment zero map for \COO{18}{2}{1} is divided by the observed one at each location where observational information is available. These ratios are point-by-point correction factors that will inform the model about where in the disk gas mass has to be added or removed. The gas profile of the initial simulation is then multiplied point-by-point by the correction factors and the resulting profile is used as an input for a second thermochemical run. We iterate over the same procedure until the relative difference between modelled and observed profiles for \COO{18}{2}{1} is $< 10\%$. During the iterations, the dust column density is kept fixed and only the gas column density is varied. Finally, we use the same gas profile to obtain synthetic observables for \CO{13}{2}{1} and \CO{12}{2}{1}. 

\begin{figure*}[!htb]
\minipage{0.32\textwidth}
  \includegraphics[width=\linewidth]{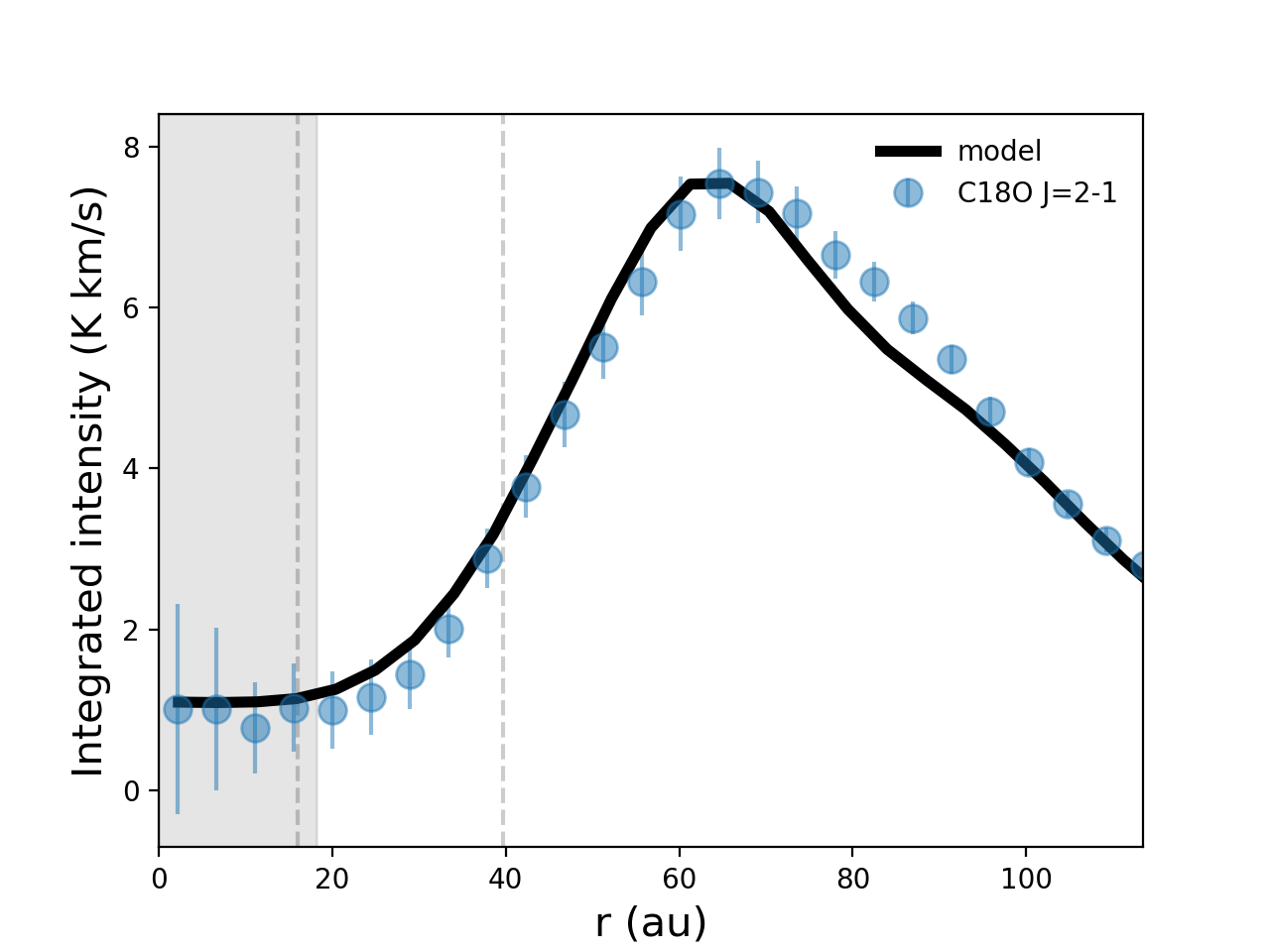}
  \label{fig:awesome_image1}
\endminipage\hfill
\minipage{0.32\textwidth}
  \includegraphics[width=\linewidth]{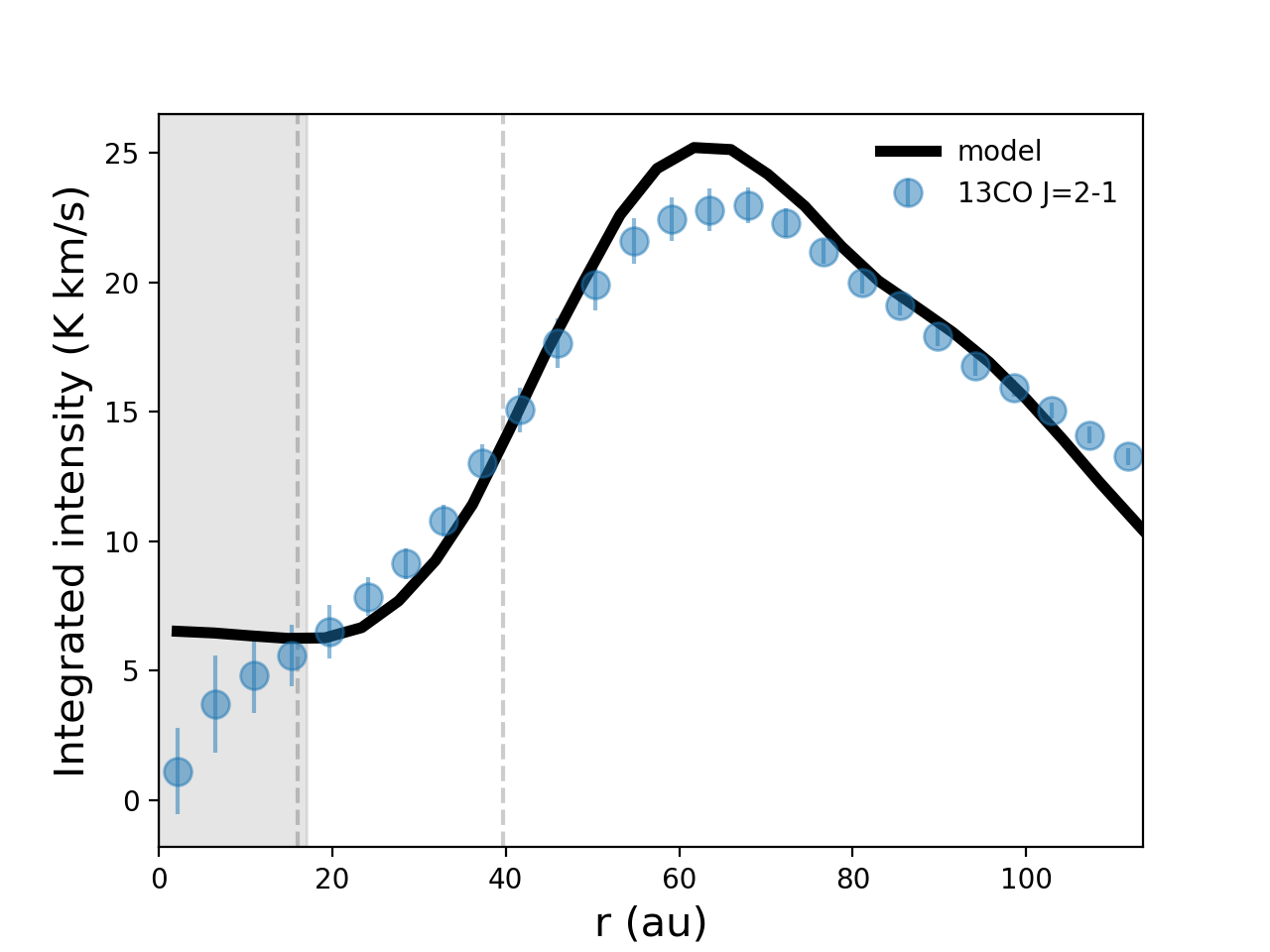}
  \label{fig:awesome_image2}
\endminipage\hfill
\minipage{0.32\textwidth}%
  \includegraphics[width=\linewidth]{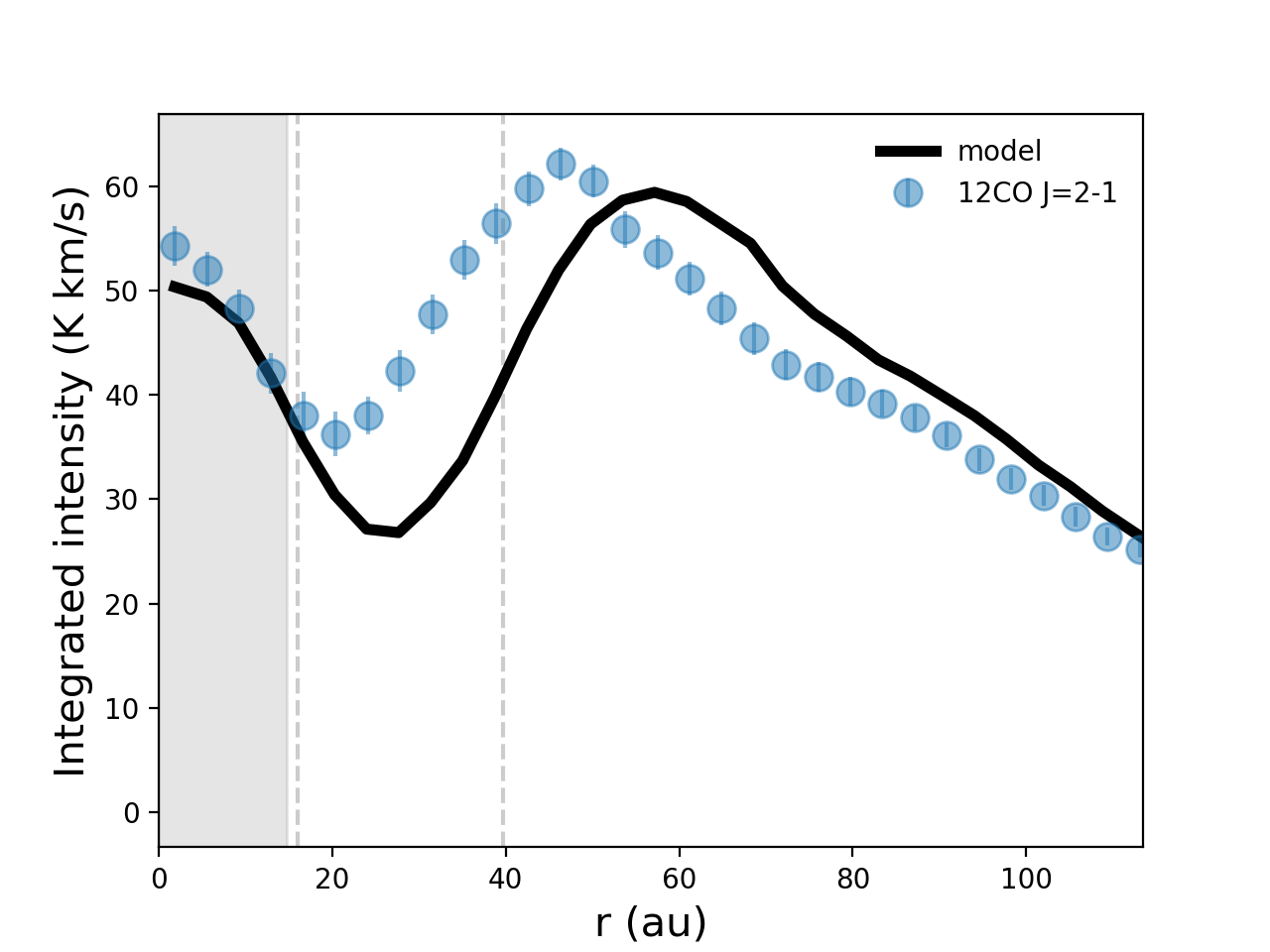}
  \label{fig:awesome_image3}
\endminipage

\caption{\small Azimuthally averaged profiles for the integrated intensity of \COO{18}{2}{1} (left), \CO{13}{2}{1} (middle) and \CO{12}{2}{1} (right). The solid lines are the modelled profiles and the blue dots are the ALMA observations. Vertical dashed lines indicate the limits of the dust gap. The shaded area indicates the resolution element of the \COO{18}{2}{1} observation of $0.16$ arcsec.}
\label{fig:model_fit}
\end{figure*}

\section{Results}
\label{sec:results}

\subsection{Benchmark for the continuum model}
\label{subsection:benchmark_continuum}
To verify the translation of the \mcmax \ disk setup into the ProDiMo 2D setup, we run the \prodimo \ model using the same constant gas-to-dust ratio of $100$ from \cite{Portilla-Revelo2022}. We simulate the continuum ALMA observation at $855\ \micron$ and the spectral energy distribution. The ray-traced image was convolved with a Gaussian beam of size $67\times 50$ mas $({\sim} 8 \times 6 \ \mathrm{au})$ and position angle of $61.5^\circ$ as in \cite{Isella2019}. A comparison to the observed data set is made in Fig. \ref{fig:sed}. Both results are in good agreement to the observations which demonstrates that the implementation of the \mcmax \ model into \prodimo \ was correctly done.   
 
Note that some fundamental differences exist between \mcmax\ and \prodimo. An important one is the treatment of the scattering. While \mcmax \ uses a full treatment of the scattering phase function, \prodimo \ only models isotropic scattering. Despite this fact, the difference in the synthetic observables for the continuum is negligible.

\subsection{Gas column density and gas-to-dust ratio profiles in the  PDS 70 disk}
\label{sect:results_profiles}

After five iterations, the procedure outlined in Sect. \ref{subsection:fit_thin} yielded a relative error $< 10\%$ with respect to the intensity profile of the \COO{18}{2}{1} observation. The fifth iteration is then the best representative model. The comparison between observed and modelled profiles both for optically thin and thick tracers is shown in Fig. \ref{fig:model_fit}. Although the gas column density is inferred based only on \coo, it does also a passable job reproducing the \CO{13}{2}{1} and \CO{12}{2}{1} data.

\begin{figure}[h!]
\centering
\includegraphics[width=1.0\linewidth]{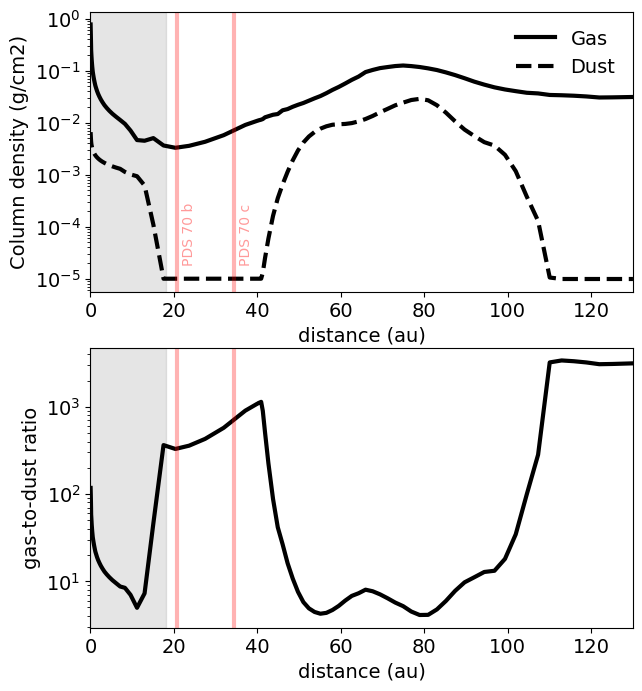}
\caption{\small (Top): predicted gas and dust column densities in $g\ \mathrm{cm}^{-2}$ versus distance to the star. (Bottom): gas-to-dust ratio. The shaded area indicates the resolution element of the \COO{18}{2}{1} observation of $0.16$ arcsec. \rt{Vertical lines indicate the location of the planets.}}
\label{fig:density_profiles}
\end{figure}

The final gas column density and gas-to-dust ratio profiles are shown in Fig. \ref{fig:density_profiles}. The value of the gas-to-dust ratio ranges from $10$ in the region beyond the cavity's outer edge ($r\gtrsim 40$ au) to almost $1000$ in the cavity. The lower value of the gas-to-dust ratio can be explained by the accumulation of dust caused by the existence of a pressure bump at the gap's outer edge. Therefore, our approach naturally points towards the expected consequence of the long-range interaction between the planets and the outer disk: the emergence of a pressure bump where dust can accumulate and grow \citep{Pinilla2012}. The higher value is driven by the low dust column density in the cavity that is expected from the observations in the continuum and their modelling. 

There is a clear modulation of the gas-to-dust profile that follows the shape of the dust column density. The gas-to-dust ratio shows strong gradients where the dust density reaches a minimum value: in the dust gap and at $r \ {\sim} 110$ au. The constant value of $10^{-5} \ \mathrm{g}\ \mathrm{cm}^{-2}$ for the dust column density in the outermost disk (see Fig. \ref{fig:density_profiles}) is also set by the ALMA observation which traces mostly (sub)millimeter-sized grains. Observations of these outer regions at shorter wavelengths are required to constrain the (sub)micron-sized population. We emphasise that the limit of applicability of our model is by construction $130$ au.

\begin{comment}
The inner disk $(r<16 \ \mathrm{au})$ seems to be depleted of gas. The gas density drops by a factor of 350 for $r<16$ au. A similar behaviour is also observed in the disk HD $169142$ where the gas density decreases by a factor of 200-500 for $r<10$ au as reported in \cite{Leemker2022}. 
\end{comment}

\subsection{Estimating the gas mass inside the dust gap}
The amount of gas inside the dust gap $\Mgas$ can be readily estimated by integrating the gas column density over the radial extension of the dust cavity from $16$ au to $41$ au and over the entire azimuth. For an axisymmetric disk, the gas mass relates to the column density via  

\begin{equation}
\label{eq:mdust}
\Mgas=\int\limits_{16\ \mathrm{au}}^{41\ \mathrm{au}} \int\limits_{0}^{2\pi} r \Sigma_{\mathrm{gas}}(r) \, \dd \phi \dd r,  
\end{equation}

\noindent which yields $\Mgas \approx 3.2 \times 10^{-3}\, \Mjup$. If we repeat this procedure for the dust component, we obtain $\Mdust \approx 5.1 \times 10^{-6}\, \Mjup$. Therefore, we find an average gas-to-dust ratio of ${\sim} 630$ within the limits of the dust gap. For the inner disk extending from $0.04$ au to $16$ au, the retrieved value for the gas mass is $7.6 \times 10^{-4}\, \Mjup$ and $7.9 \times 10^{-5}\, \Mjup$ for the dust mass, implying a gas-to-dust ratio of ${\sim} 10$. These quantities for the inner disk must be taken with caution due to the limited spatial resolution of our data set.   

The total gas mass of the best representative model is $>3.0 \times 10^{-4} \, \Msun$ (inside $130$ au). The radial extent of the gas disk in PDS 70 is expected to go well beyond that traced by the (sub)millimeter dust. Thus, we interpret this value as a lower limit.  

\section{Discussion}
\label{sec:discusion}

\subsection{Linking gas depletion to planet masses}
\label{sec:depletion_and_mass}
Models of planet formation predict the opening of a gap in the gas distribution by planets of mass $>15 \, \Mearth$. For a disk hosting a single gap-carving planet, the mass of the planet $\Mp$ correlates with the depth of the gap. \cite{Kanagawa2015} showed that this relation is given by: 

\begin{equation}
\label{eq:kanawaga15}
\frac{\Mp}{\Ms}=\Bigg[25\ \bigg(\frac{1}{\Sigmax{p}/\Sigmax{0}}-1\bigg)\ \hp^5\ \alpha \Bigg]^{1/2},    
\end{equation}

\noindent where $\Sigmax{p}$ and $\Sigmax{0}$ are gas column densities measured at the bottom of the cavity and at the gap's edge, respectively. Their ratio represents the level of gas depletion in the cavity. Equation (\ref{eq:kanawaga15}) also depends on the scale height and turbulence strength measured by the Shakura-Sunyaev $\alpha$ parameter \citep{Shakura1973}. For fixed values of the scale height and turbulence parameter, it is clear that 

\begin{equation}
\label{eq:kanawaga15_short}
    \frac{\Mp}{\Ms} \propto \ \bigg(\frac{\Sigmax{p}}{\Sigmax{0}}\bigg)^{-1/2};
\end{equation}

\noindent the mass of a planet scales with the inverse of the square root of the gas depletion it causes. 

For the multiple-planet case, \cite{Duffell2015} provide a set of closed fitting formulas relating the level of gas depletion to the planet mass (see Eq. (14) in that paper). Planets were assumed to have the same mass and to move on circular orbits under the effect of the combined gravitational potential of the star and the other planets. They were located in a 2:1 mean motion resonance with one another. The star was at rest and unperturbed by the planets. The disk was assumed to be globally isothermal, with a constant sound speed that satisfies a scale height value of $0.05$ at the location of the outermost planet. The functional form of the fitting formula depends on how the mass of the planet is related to a critical value of the planet-to-star mass ratio ($q_\mathrm{crit}$) which is readily derived from the gap merging criterion, i.e. from the commensurability of the minimum separation between the two planets to the radial extent of their spheres of influence. Using the best fit orbital solution for PDS 70 b and c from \cite{Wang2021}, we found $q_\mathrm{crit}=0.002$. Particularly, for planet-to-stellar mass ratio values $\gtrsim 2.2\times q_\mathrm{crit}$, the scaling law from \cite{Duffell2015} takes the form: 

\begin{equation}
\label{eq:duffell15}
    \frac{\Mp}{\Ms} \propto \ \bigg(\frac{\Sigmax{p}}{\Sigmax{0}}\bigg)^{-1/4}.
\end{equation}

From the scaling relations given by Eqs. (\ref{eq:kanawaga15_short}) and (\ref{eq:duffell15}) we see that in order to explain the same level of gas depletion, a disk with multiple planets would require them to be more massive than the single-planet counterpart.

Following the theoretical results by \cite{Duffell2015} and \cite{Kanagawa2015}, we use the results in Fig. \ref{fig:density_profiles} to estimate the degree of gas depletion in the gap of the PDS 70 disk. The location of the gap edge for the gas distribution is assumed to match that of the peak value and therefore we adopt $\Sigmax{0}=0.13\, \mathrm{g}\, \mathrm{cm}^{-2}$ at $r=75 \ \mathrm{au}$. The minimum occurs at $20.4$ au and has a value of $\Sigma_\mathrm{min}=3.3\times 10^{-3} \ \mathrm{g}\ \mathrm{cm}^{-2}$. Interestingly, the location of the minimum matches the best fit solution for the semi-major axis of PDS 70 b found by \cite{Wang2021}. These values imply a gas density drop by a factor of ${\sim} 39$ at $20$ au. \rt{In order to be consistent with the prescription in \cite{Duffell2015} for the multiple-planet case, the gas depletion must be re-defined with respect to the value measured at the location of the outermost planet.} The semi-major axis of planet c is $a_\mathrm{c}=34.3$ au \citep{Wang2021} and at that radius, the gas surface density is $\Sigmax{p}=7.0\times 10^{-3} \ \mathrm{g}\ \mathrm{cm}^{-2}$. Therefore, the level of gas depletion is:

\begin{equation}
\label{eq:gas_depletion_level}
    \frac{\Sigmax{p}}{\Sigmax{0}}\approx 0.054,
\end{equation}

\noindent which is equivalent to a density drop factor of $\mathbf{{\sim} 19}$ at $a_\mathrm{c}=34$ au.

\begin{figure}[h!]
\centering
\includegraphics[width=1.0\linewidth]{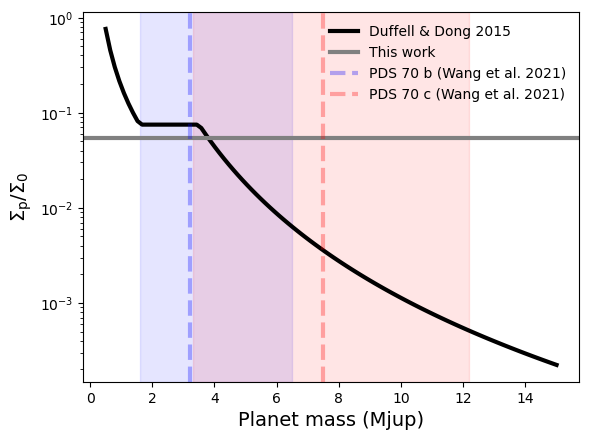}
\caption{\small The value for the gas depletion inferred from our model (grey line) is overlaid to the scaling relation from \cite{Duffell2015} (black line). The two curves intersect at $\sim 4\, \Mjup$. Dashed lines and shaded areas indicate the mean value and the $68\%$ confidence interval for the mass of PDS 70 b and c given by \cite{Wang2021}.}
\label{fig:planet_mass}
\end{figure}

We directly compare Eq. (\ref{eq:gas_depletion_level}) with Eq.(14) from \cite{Duffell2015} which is the fitting formula connecting the gas depletion with the planet mass. The result is shown in Fig. \ref{fig:planet_mass} where the horizontal line intersects the theoretical curve at a planet mass of $4 \, \Mjup$. We stress that the planet mass derived through this method is degenerate by construction because the underlying hydrodynamical models assumed equal mass planets. Accordingly, we constrain the mass of each planet to be roughly $4 \Mjup$. For comparison, we overlay the mass estimates for both planets given by \cite{Wang2021} which were obtained via analysis of near-infrared data. The respective shaded areas indicate the $68\%$ confidence interval in their mass determination. In conclusion, the value that we predict for the mass of each planet in the PDS 70 disk via analysis of submillimeter data is in good agreement with literature values based on infrared techniques\footnote{Planetary masses found in \cite{Wang2021} are: $M_\mathrm{b}=3.2_{-1.6}^{+3.3} \, \Mjup$ and $M_\mathrm{c}=7.5_{-4.2}^{+4.7} \, \Mjup$. Superscript and subscript represent the 68\% credible interval.}. 

Although our $4 \, \Mjup$ estimate is close to the mean value for PDS 70 b according to \cite{Wang2021}, it is more to the lower side for PDS 70 c. For the latter case, special considerations have to be made to account for obscuration effects of the circumplanetary disk (CPD) and, to a lesser degree, of the circumstellar outer disk. As for the CPD effect, we can crudely approximate the fraction of the planet mass that the CPD would represent. In \cite{Portilla-Revelo2022}, we estimated the CPD dust mass to be ${\sim} 0.1 \Mearth$. Assuming a gas-to-dust ratio of $100$ for the CPD, its total mass would account for nearly $1\%$ of the estimated planet mass. Therefore, although our method is agnostic to the presence of a CPD, we do not expect that to impact our results.

%\subsubsection{Dependence on model parameters}
Although \cite{Duffell2015} fixed the values of the turbulence strength and the pressure scale height in their simulations, we expect the mass estimation in the multi-planet case to depend on those parameters as well, just like in \cite{Kanagawa2015} for the single-planet case (see Eq. \ref{eq:kanawaga15}). Due to the forward modelling approach implemented in this work, we cannot provide a full statistical account of the uncertainties associated to each parameter in the thermo-chemical model. Instead, in the following subsections we investigate the robustness of our mass estimate for some key parameters. This is not intended to be an exhaustive exploration of the whole range of values those quantities can adopt but rather limited to a series of physically motivated scenarios. 

\subsubsection{The effect of dust filtration on the pressure scale height}
\label{sec:filtration}
Simulations in \cite{Duffell2015} and \cite{Kanagawa2015} did not account for particle filtration at the outer edge of the cavity. In other words, they assumed that the dust grains, even the mm-sized ones, follow the same radial distribution as the sub-$\micron$ grains and the gas particles. As dust is the main source of opacity at visual and infrared wavelengths, the way how it is distributed can alter the thermal structure of the disk and the gas physics. Many theoretical studies suggest that the gas and dust distributions can largely deviate from each other when substructures like gaps opened by planets are present \citep{Rice2006}. This phenomena might be in operation in PDS 70 as suggested by the hydrodynamical simulations from \cite{Bae2019} where only sub-micron sized grains were observed to populate the inner disk after $0.6$ Myr. A similar scenario is supported by the SED fitting and scattered-light modelling of the same source in \cite{Portilla-Revelo2022}. However, as demonstrated in Sect. \ref{subsection:benchmark_continuum}, the multiple degeneracies among the parameters of a forward model leave room to explain the same observables with a homogeneous dust size distribution. In principle, the scenario of a disk with segregated grains can be conciliated with the non-segregated case by claiming efficient grain growth in the inner disk but inferring the presence of this process in PDS 70 would require a dedicated modelling including the appropriate physics. 

Instead, we estimate the effect that dust filtration could have on our derived planet masses. Dust filtration alters the opacity properties in the inner zone. For a fixed dust column density, an inner disk ($r\lesssim 16$ au) populated with grains of maximum size $\amax \ {\sim} 100 \, \micron$ is expected to have a higher visual and infrared opacity than the outer disk where $\amax \ {\sim} 3$ mm. This affects the thermal structure of the disk which in turn will impact the scale height via the gas temperature. This will affect the planet mass estimate according to Eq. (\ref{eq:kanawaga15}) 

Assuming LTE conditions and that the gas and dust temperatures are equal, which is a good approximation in the midplane where the planets reside, it is easy to make a semi-analytical study of the impact of dust filtration on the planet mass estimate. To achieve this goal, we first compute the dust temperature for the two size distributions we are interested in. Assuming a balance between stellar heating and black-body radiation cooling, the dust equilibrium temperature $T_\mathrm{d}$ is given by the solution to the implicit equation 

\begin{equation}
\label{eq:Td_eq}
    T_\mathrm{d}^4 = \bigg(\frac{R_*}{2r}\bigg)^2 \frac{\kabsave_*}{\kabsave_{T_\mathrm{d}}}\ T_*^4,
\end{equation}

\noindent where $R_*$ and $T_*$ are the stellar radius and effective temperature, $r$ is the distance from the planet to the star, and the absorption opacities $\kabs$ are averaged over the stellar spectrum in the numerator, and over the black-body spectrum in the denominator. We solve Eq. (\ref{eq:Td_eq}) for two values of the maximum grain size in the inner zone $\amax=100\ \micron$ and $\amax=3$ mm, which represent respectively two scenarios when filtration is present at the gap's edge and when it is not. The calculation yields $T_\mathrm{d,100\, \micron} / T_\mathrm{d,3\, \mathrm{mm}} = 1.2$. Finally, let us assume that the mass in the multiple-planet case also scales as $\hp^{5/2}$ as in the single planet case (Eq. \ref{eq:kanawaga15}). Due to the square-root dependence of the sound speed on the temperature\footnote{Isothermal sound speed is defined as $\displaystyle c_\mathrm{s}^2 = \frac{k_\mathrm{B} T}{\mu m_\mathrm{H}}$, where $m_\mathrm{H}$ and $\mu$ are the mass of the proton and the mean molecular weight of a gas of cosmic composition, respectively. Scale height and sound speed are related via $\displaystyle h_\mathrm{p} = \frac{c_\mathrm{p}}{r_\mathrm{p} \Omega_\mathrm{p}}$ where $\Omega_\mathrm{p}$ is the Keplerian frequency at distance $r_\mathrm{p}$.}, this difference of $20\%$ in temperature translates into a difference of just $10\%$ in the sound speed. Accordingly, the relative difference on the scale heights should be of the order of $10\%$ as well. Finally, assuming that all the parameters but the scale height in Eq. (\ref{eq:kanawaga15}) remain constant, we expect a difference of $26 \%$ in the the planet mass. This brings our initial estimate of $\mathbf{4\ \Mjup}$ up to $\mathbf{5 \, \Mjup}$.          

\subsubsection{Dependence on $\alpha$ and $\Sigmax{0}$}
From Eq.(\ref{eq:kanawaga15}), it follows that the planet mass depends on both the Shakura-Sunyaev $\alpha$ parameter and the unperturbed $\Sigmax{0}$ density to the power one-half (to the power one-fourth in the multiple-planet case). The values of those parameters are model dependent and in principle correlated and so are their uncertainties. However that correlation is not trivial to establish because of the several intricacies among the heating and cooling mechanisms and the chemical processes which in the end determine the shape of the gas density. The simplest method to circumvent this problem is to work on an alternative forward model that uses a different value of $\alpha$ to derive a new best fit gas density profile. 

Building on the dust model in \cite{Portilla-Revelo2022} that used $\alpha=2\times 10^{-2}$ and $\amax=100\ \micron$, and $\alpha=10^{-2}$ and $\amax=3$ mm, in the inner and outer zones respectively, we construct an alternative model for the gas that reproduces the \COO{18}{2}{1} observation. The retrieved density profile shows a peak value of $0.17 \ \mathrm{g}\ \mathrm{cm}^{-2}$ at $76$ au while the density at the location of PDS 70 c is $7.4\times 10^{-3} \ \mathrm{g}\ \mathrm{cm}^{-2}$. These values are very similar to those presented in Sect. \ref{sec:depletion_and_mass} for the best fit model and therefore, following \cite{Duffell2015} for the multiple-planet case, the planet masses are expected to be similar too. 

Conversely for the single-planet case, we expect from Eq. (\ref{eq:kanawaga15}) a planet twice as massive due to the factor of four in the ratio of $\alpha$ parameters in both models. However, the alternative model also predicts a hotter midplane which is in part explained by the higher continuum opacity of the inner zone. In fact, the representative model in Sect. \ref{subsection:fit_thin} predicts $T_\mathrm{gas}=16$ K for the gas temperature at the location of PDS 70 c whereas the alternative model has $T_\mathrm{gas}=21$ K. Introducing these dependencies into Eq. (\ref{eq:kanawaga15}), we find that the planet mass increases to $9 \ \Mjup$, a factor of $2.4$ difference in the mass estimate. This high value still falls within the predicted mass of PDS 70 c but is beyond that of planet b. 

\rt{Although these experiments help to quantify the impact of $\alpha$ on the mass estimate, they do not break the degeneracy between both quantities, which is a more fundamental issue that should be addressed with an expanded parameter space exploration in the hydrodynamical simulations. Note also that the $\alpha$ parameter used in the hydrodynamical models and that used in \prodimo \ act in different ways, i.e. driving gas flows and determining vertical settling of dust grains, respectively. The viscosity responsible for these two phenomena might not have the same physical origin. In the future, the two modelling approaches should be directly coupled in an attempt to improve the method. Additionally, our forward modelling suggests that $\alpha=5\times 10^{-3}$ but this by no means implies that we can rule out lower values. An enhanced settling could bring $\Sigmax{0}$ to lower values such that the depletion level coincides with the plateau between $q_\mathrm{crit}$ and $2.2\times q_\mathrm{crit}$ seen in Fig. \ref{fig:planet_mass} where the mass is entirely unconstrained. Therefore, an independent measurement of $\alpha$ in PDS 70 is needed and should be the aim of a future work.} 

The results presented in this section show that although they strongly depend on the model, our approach is capable of constraining the mass of embedded planets within a factor of a few for reasonable choices in the disk's parameters. Since those parameters are dictated by a dedicated modelling of multiple observables, they are prone to be refined as new observations come along. We demonstrated here that such an approach has the potential to naturally reduce the uncertainties in the mass estimation of young protoplanets, especially in those cases where direct observations prove impossible.  

\subsection{Gas stirring in the gap of PDS 70}
\label{sec:gas_stirring}
\rt{An upper limit for the level of gas depletion can be obtained applying Eq. (\ref{eq:kanawaga15}) to the lowest value of the combined planet mass that is consistent with the VLTI/GRAVITY measurements \citep{Wang2021}. Assume this lowest value equals the arithmetic mean of the values at the lower end in the $68\%$ credible intervals in \citealt{Wang2021}. This is $2.5\ \Mjup$. Solving Eq. (\ref{eq:kanawaga15}) for the level of gas depletion, we find the upper limit to be\footnote{\rt{In this step we used $\alpha=5\times 10^{-3}$. From the model, we also find a sound speed value of $0.26 \ \mathrm{km}\ \mathrm{s}^{-1}$ at $34.3$ au, implying $h_\mathrm{p}=0.059$ at the same radius.}} $9.0\times 10^{-3}$ which is a factor of $6$ lower than our estimate given in Eq. \ref{eq:gas_depletion_level} (note that this factor will only increase if we pick the mean planet masses or those at the upper end of the interval). To conciliate both values, we would have to assume a lower combined mass, but this would be inconsistent with the constraints imposed by the infrared data. Another way to bring these factors close to each other is invoking gas stirring due to the perturbative effect that each planet exerts on the gas in the common gap. In fact, Figure 3 in \cite{Duffell2015} predicts that gas stirring due to two or more planets can naturally explain gas gaps being shallower by at least a factor of $10$ compared to the gap depths expected in the single planet case.}

\rt{We stress again that this analysis depends strongly on the ability of our model to capture the actual physical state of the disk. Under that assumption, our results support the idea that gas stirring processes are at play in the planet-hosting gap in the PDS 70 disk.}

\subsection{Modelling optically thick lines}
\label{sec:fit_thick}
The abundance of \cooo \ is almost two orders of magnitude larger than the less common and moderately optically thick isotopologue \co. Hence, we assumed the former to be optically thick. Unlike the thin case, optically thick emission is mainly driven by temperature.

First, we can compare our results to the line flux emitted by the disk within the first $130$ au. This region is delimited by the red dashed ellipse in Fig \ref{fig:M0_compara}. The model predicts an integrated value of $3624 \ \mathrm{mJy}\ \mathrm{km}\ \mathrm{s}^{-1}$ for \CO{12}{2}{1}. This is close to the observed value of $3545 \ \mathrm{mJy}\ \mathrm{km}\ \mathrm{s}^{-1}$. For \CO{13}{2}{1} we obtain a value of $1355 \ \mathrm{mJy}\ \mathrm{km}\ \mathrm{s}^{-1}$ which is again in good agreement to the observed value of $1331 \ \mathrm{mJy}\ \mathrm{km}\ \mathrm{s}^{-1}$.

\begin{figure}[h!]
\centering
\includegraphics[width=1.0\linewidth]{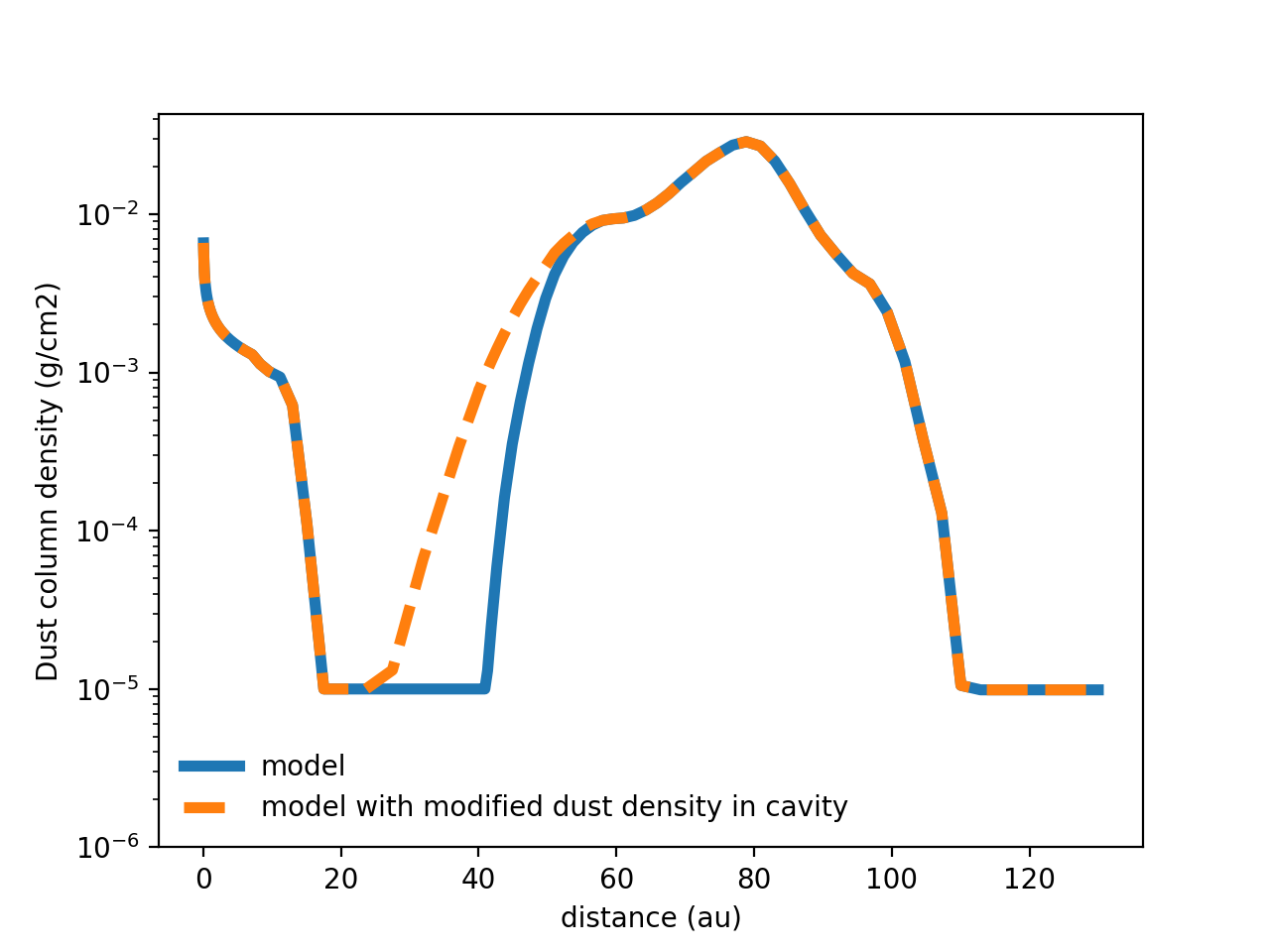}
\caption{\small Dust column density profile for the PDS 70 disk (solid line, same as in Fig. \ref{fig:density_profiles}) and that from a similar disk with $30\%$ extra dust mass between $27$ and $60$ au.}
\label{fig:dust_profile_modified}
\end{figure}

\begin{figure*}[!htb]

\minipage{0.5\textwidth}
  \includegraphics[width=\linewidth]{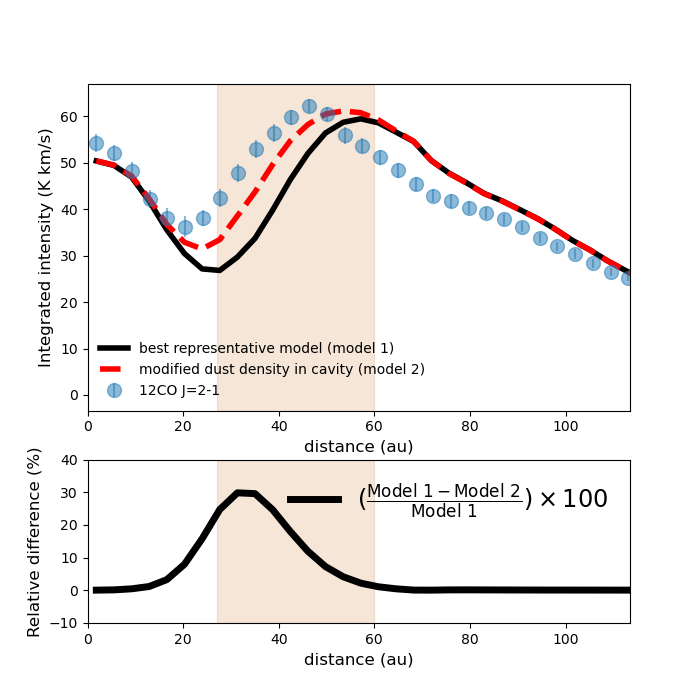}
  \label{fig:awesome_image1}
\endminipage\hfill
\minipage{0.5\textwidth}
  \includegraphics[width=\linewidth]{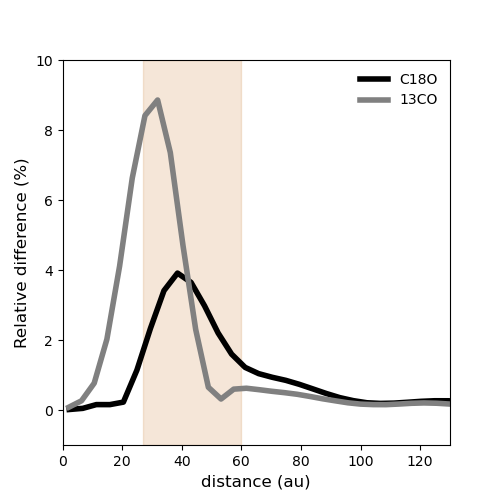}
  \label{fig:awesome_image2}
\endminipage

\caption{\small (Top left): Azimuthal profiles of the \CO{12}{2}{1} integrated maps from two models with slightly different dust distribution between $27$ au and $60$ au. Blue dots are the observations. The shaded area indicates the region where the dust mass is artificially increased by $30\%$ to boost heating by dust thermal accommodation. (Bottom left): The relative difference of modelled intensities with respect to the best representative model for the \CO{12}{2}{1} case. (Right): Relative difference of integrated intensities for the \CO{13}{2}{1} and \COO{18}{2}{1} cases.}
\label{fig:temperature_effect}
\end{figure*}

Now we analyse the spatially resolved intensities. The right panel in Fig. \ref{fig:model_fit} compares the simulated \cooo \ to its observational counterpart. The maximum deviation from the observation happens within the limits of the dust gap, specially near the outer edge, whereas outside the gap the fits seems passable. This suggest that although several parameters can be tweaked to alleviate the discrepancy, we should focus our attention on those that are expected to have a local effect in the gas temperature. Analysis of the heating mechanisms contributing to the gas temperature shows that inside the gap, for values of $z/r \ {\sim} 0.1$, heating by $\mathrm{H}_2$ formation on dust is the leading heating mechanism followed by dust thermal accommodation. Both of those heating mechanisms are proportional to the dust density. Therefore we modify the dust distribution keeping constant the gas density and explore the impact on the \cooo \ emission. We artificially increased the dust mass contained between $27$ and $60$ au by a factor of ${\sim} 30\%$ by deforming the column density using a Gaussian function with a standard deviation of $9$ au within those limits (Fig. \ref{fig:dust_profile_modified}). We run the thermochemical model on top of this dust profile and the radial profile predicted for the \cooo \ emission is shown in Fig. (\ref{fig:temperature_effect}). As expected, the higher column density near the gap's outer edge has a localised effect on the \cooo \ intensity which now agrees better with the observation. The modified dust density guarantees that the fit to the other observables is not severely undermined.

In fact, the left panel of Fig. \ref{fig:temperature_effect} shows the radial profiles of the \CO{12}{2}{1} moment zero maps from the best representative model and from a similar model that used the modified dust surface density of Fig. \ref{fig:dust_profile_modified}. The predicted intensities differ only in the region where the dust density was increased and the maximum difference is $30\%$. For $\co$ and $\coo$, the relative deviations are minor and as low as $4\%$ for the \coo \ case (Fig. \ref{fig:temperature_effect}, right panel). We demonstrated with this experiment that there is a subset of the parameter space that can be tweaked to bring the optically thick emission closer to the observation keeping the gas density and the level of turbulence unchanged. It also shows that these modifications of the temperature structure will not change substantially our main conclusions driven by the analysis of optically thin emission.

Modifying the dust profile is by no means the only way to improve the fit to thick lines. In recent papers, it has been shown for other disks, that gas cavities are smaller than the dust counterpart and that the temperatures in the cavities tend to increase (see e.g. \citealt{Wolfer2023} and \citealt{Leemker2022}). If the same applies to PDS 70, these effects combined can bring the \cooo \ profiles into closer agreement with the observations without requiring to modify the dust distribution.

Polycyclic aromatic hydrocarbons (PAHs) might have an influence in setting the gas temperature within gaps and cavities \citep{Leemker2022}. For example, \cite{Calahan2021} found for HD 163296 the PAHs have a significant contribution to the gas heating only in those layers where gas and dust temperatures decouple. Given the lack of constraints on the PAH abundance in T Tauri disks \citep{Geers2006}, we did not perform a detailed exploration for this parameter and we use a fixed abundance of $1\%$ of the ISM value. 

The thermal accommodation coefficient that regulates the energy exchange rate in inelastic collisions between gas and dust particles \citep{Burke1983} and can take any value between 0 and 1, was not explored in detail and we used a fixed value of $0.2$. However, it is clear that a single value for this quantity will be insufficient if the dust properties change throughout the disk.

\subsection{Emitting surfaces}
\label{sec:esurface}
The fact that molecular line emission originates well above the midplane sets important constraints on the modelling, mainly when dealing with optically thick transitions. From the \prodimo \ models it is straightforward to extract information about where in the disk radiation from a certain molecular species is being emitted. We define here the line emitting region as the area from which the $15\%-85\%$ percentile of the radiation is being emitted along the radial and vertical directions. The comparison between the area of emission to the emitting surface observationally derived can be used as another proof of the robustness of the model. 

\begin{figure}[h!]
\centering
\includegraphics[width=1.0\linewidth]{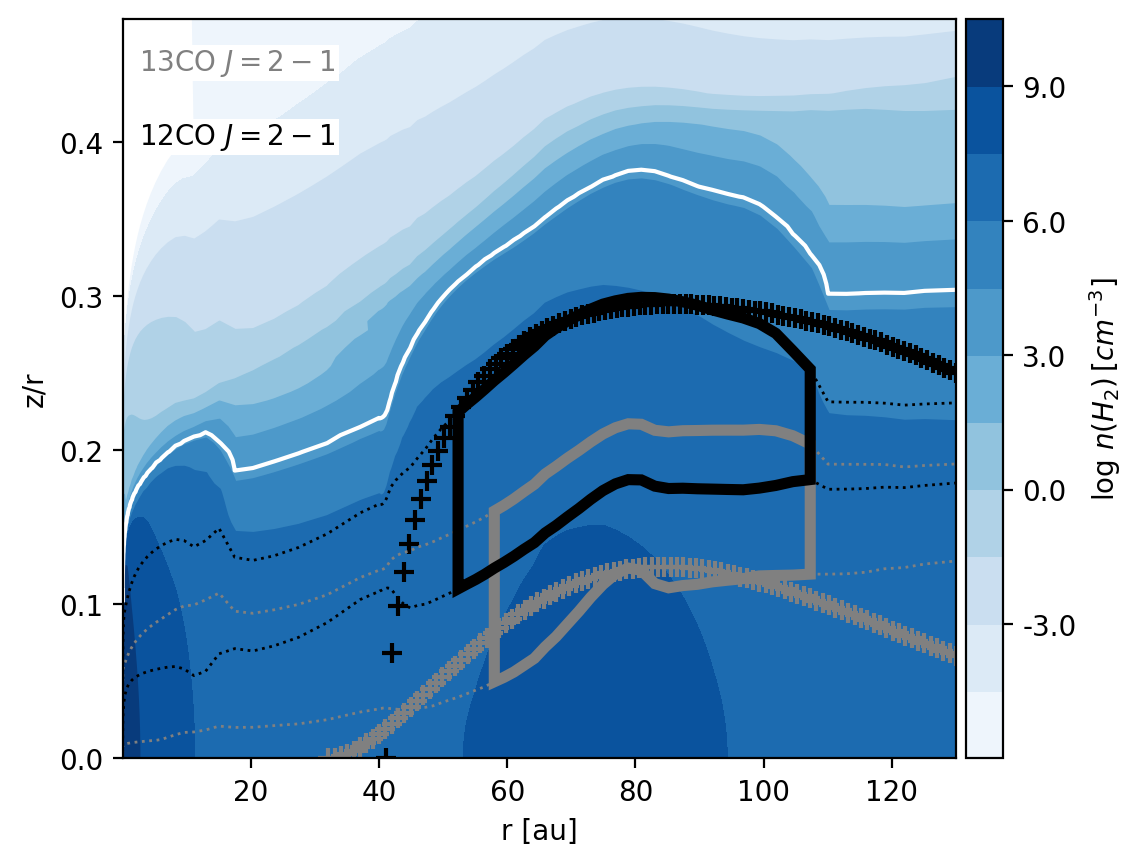}
\caption{\small Emitting regions from which $15\% \ \mathrm{to} \ 85\%$ of the \CO{13}{2}{1} (grey solid box) and \CO{12}{2}{1} (black solid box) radiation is emitted. Dotted lines indicate the emitting region in the vertical direction only. The grey and black crosses are mapped from the fitting functions to the respective emitting surfaces derived in Law et al. in prep. The white curve is the $1.5 \times 10^4 \ \mathrm{cm}^{-3}$ contour for the critical density of the \CO{12}{2}{1} transition with $\mathrm{H}_2$ as the collision partner. The color map is the gas density.}
\label{fig:emitting_region}
\end{figure}

Law et al. (in prep) derive line emitting surfaces for several molecules in PDS 70. We use the parameters of the best fit solution in that paper to compare with the line emitting regions for \co \ and \cooo \  predicted by the model (see Fig. \ref{fig:emitting_region}). In both cases, the emitting area is below the critical density threshold for \CO{12}{2}{1} given by the white curve, justifying the use of LTE excitation. 

In general, there is a good agreement between the area of emission and the emitting surface from the observation inside $100$ au, but at larger distances the model tends to overpredict the height of emission of \co \ while it underpredicts that of \cooo. An extension of the model to larger semi-major axes is needed to study the behaviour of the emitting regions in the outskirts of the disk. 

\subsection{Variations in the C/O ratio}
It has been shown that the abundance of CO is correlated to the C/O ratio, with high C/O ratios implying lower CO abundances \citep{Bergin2016}.  
Combining high resolution ALMA observations and dedicated thermochemical models for the AS 209, HD 163296 and MWC 480 disks, \cite{Bosman2021} showed that a high C/O ratio $(>2)$ together with a strong depletion of small dust grains are needed to reproduce the measured peak values of the $\mathrm{C}_2\mathrm{H}$ column densities.

Our best representative model uses a canonical solar C/O ratio of $0.457$ that is assumed to be constant across the disk. To estimate the effect that a potentially higher C/O ratio would have in our synthetic observables, we decreased the elemental abundance of oxygen such that the C/O ratio goes up to ${\sim}1.1$. We use this value to make sure that the chemistry switches into the regime of carbon chemistry (CO formation locks up all oxygen). We ran a simulation keeping constant the rest of the parameters and the results are shown in Fig. (\ref{fig:C2O_variation}). The quality of the fit to the brightness profiles is not severely diminished and an additional fine-tune to the total gas distribution seems unnecessary. On the basis of this simple experiment, we do not expect substantial changes in the synthetic CO observables due to a variation of the C/O ratio slightly above the unity.

%A comprehensive study of the impact of C/O variations on the inferred gas distribution requires testing the model against other observables. The forward model developed for this work can be used as a starting point for such a study.  

\section{Conclusions}
\label{sec:conclusions}
We studied the gas distribution of the PDS 70 disk using radiative transfer modelling of dust and line emission with a special focus on the dust depleted gap where the two planets reside. We use a subset of the chemical inventory from \cite{Facchini2021} to develop a thermochemical model that explains ALMA band 6 observations of \COO{18}{2}{1}, \CO{13}{2}{1}, and \CO{12}{2}{1}. The continuum emission at $855 \ \micron$ and the photometry towards the source are also reproduced. \rt{The modelling results are used to characterise the spatial distribution of gas and dust in the disk, and the results are interpreted in the light of the theory of planet-disk interaction.} Our main conclusions are: 

\begin{enumerate}
\item We find a position-dependent gas-to-dust ratio profile with values ranging from ${\sim} 10 - 1000$ across the first $130$ au in the disk. The profile depicts a sharp transition from the gap to the outer disk. This is consistent with the existence a pressure maxima outside the orbit of the planets originated by the dynamical interaction between the planets and the outer disk.     

\item The model suggests a gas mass of $3.2 \times 10^{-3} \, \Mjup$ within the dust depleted gap where two planets reside. Combined with the dust mass constrained from modelling of the continuum emission, the spatially averaged gas-to-dust ratio in the gap is ${\sim} 630$.   

\item \rt{Combining the level of gas depletion in the gap with the results of hydrodynamical models by \cite{Duffell2015} we estimate a mass value of $4 \ \Mjup$ for each of the two planets detected in the gap of the PDS 70 disk. This value is consistent with independent constraints based on infrared spectroscopy.} We emphasise that the mass here derived is equal for both planets since this is an inherited limitation from the underlying hydrodynamical model. Although the planet masses estimated here depend on the model parameters, specially on the turbulence strength, we demonstrate that the values are robust within a factor of a few for reasonable choices in the model parameters. This work demonstrates the power of using a multi-wavelength and synergistic approach to characterise planets in the embedded phase.

\item \rt{Based on the best representative model, we find a level of gas depletion that is lower by a factor of $6$ with respect to the value expected from the action of a single planet. This could be a result of the gas stirring produced in the gap by the combined gravitational effect of the two planets.}

\item We estimate that $85\%$ of the \CO{12}{2}{1} and \CO{13}{2}{1} emission originating in the outer disk ($r>60$ au) comes from of $z > 6$ au and $z > 3$ au above the midplane, respectively.  
\end{enumerate}

In this work, we demonstrated the capabilities that forward thermo-chemical modelling coupled to well-established results from hydrodynamical models have to exploit a relatively new parameter space of planet formation theories: direct observations of young planets embedded in their progenitor disk. As new cases of study emerge, multiwavelength and multiphase observations at high resolutions and sensitivities will be essential. This motivates the development of new, or the refinement of existing models, that can help to characterise not only the natal disk but also the nascent planets.

\begin{acknowledgements}
We thank the anonymous referee for a constructive report and for suggesting the analysis that inspired Sect. \ref{sec:gas_stirring}. We thank prof. Paul Duffell for a prompt response to our inquiry on the definition of the gas depletion in the gap. This work is partly supported by the Netherlands Research School for Astronomy (NOVA). S.F. is funded by the European Union under the European Union’s Horizon Europe Research \& Innovation Programme 101076613 (UNVEIL). CHR acknowledges the support of the Deutsche Forschungsgemeinschaft (DFG, German Research Foundation) Research Unit “Transition
discs” - 325594231. This research was supported by the Excellence Cluster ORIGINS which is funded by the Deutsche Forschungsgemeinschaft (DFG, German Research Foundation) under Germany’s Excellence Strategy - EXC-2094 - 390783311. CHR is grateful for support from the Max Planck Society. This project has received funding from the European Research Council (ERC) under the European Union's Horizon 2020 research and innovation programme (PROTOPLANETS, grant agreement No. {$\sim$}101002188).
%We thank the anonymous referee for a constructive report that improved the quality of the paper. This work is partly supported by the Netherlands Research School for Astronomy (NOVA). CHR acknowledges support from the DFG Research Unit "Transition Disks" project number 325594231 (FOR 2634/1 and FOR 2634/2). CHR is grateful for support from the Max Planck Society.  
\end{acknowledgements}

\bibliographystyle{aa}
\bibliography{references}

\begin{thebibliography}{55}
\expandafter\ifx\csname natexlab\endcsname\relax\def\natexlab#1{#1}\fi

\bibitem[{{Bae} {et~al.}(2019){Bae}, {Zhu}, {Baruteau}, {Benisty}, {Dullemond},
  {Facchini}, {Isella}, {Keppler}, {P{\'e}rez}, \& {Teague}}]{Bae2019}
{Bae}, J., {Zhu}, Z., {Baruteau}, C., {et~al.} 2019, \apjl, 884, L41

\bibitem[{{Benisty} {et~al.}(2022){Benisty}, {Dominik}, {Follette}, {Garufi},
  {Ginski}, {Hashimoto}, {Keppler}, {Kley}, \& {Monnier}}]{Benisty2022}
{Benisty}, M., {Dominik}, C., {Follette}, K., {et~al.} 2022, arXiv e-prints,
  arXiv:2203.09991

\bibitem[{{Bergin} {et~al.}(2016){Bergin}, {Du}, {Cleeves}, {Blake}, {Schwarz},
  {Visser}, \& {Zhang}}]{Bergin2016}
{Bergin}, E.~A., {Du}, F., {Cleeves}, L.~I., {et~al.} 2016, \apj, 831, 101

\bibitem[{{Bosman} {et~al.}(2021){Bosman}, {Alarc{\'o}n}, {Bergin}, {Zhang},
  {van't Hoff}, {{\"O}berg}, {Guzm{\'a}n}, {Walsh}, {Aikawa}, {Andrews},
  {Bergner}, {Booth}, {Cataldi}, {Cleeves}, {Czekala}, {Furuya}, {Huang},
  {Ilee}, {Law}, {Le Gal}, {Liu}, {Long}, {Loomis}, {M{\'e}nard}, {Nomura},
  {Qi}, {Schwarz}, {Teague}, {Tsukagoshi}, {Yamato}, \& {Wilner}}]{Bosman2021}
{Bosman}, A.~D., {Alarc{\'o}n}, F., {Bergin}, E.~A., {et~al.} 2021, \apjs, 257,
  7

\bibitem[{{Burke} \& {Hollenbach}(1983)}]{Burke1983}
{Burke}, J.~R. \& {Hollenbach}, D.~J. 1983, \apj, 265, 223

\bibitem[{{Calahan} {et~al.}(2021){Calahan}, {Bergin}, {Zhang}, {Schwarz},
  {{\"O}berg}, {Guzm{\'a}n}, {Walsh}, {Aikawa}, {Alarc{\'o}n}, {Andrews},
  {Bae}, {Bergner}, {Booth}, {Bosman}, {Cataldi}, {Czekala}, {Huang}, {Ilee},
  {Law}, {Le Gal}, {Long}, {Loomis}, {M{\'e}nard}, {Nomura}, {Qi}, {Teague},
  {van't Hoff}, {Wilner}, \& {Yamato}}]{Calahan2021}
{Calahan}, J.~K., {Bergin}, E.~A., {Zhang}, K., {et~al.} 2021, \apjs, 257, 17

\bibitem[{{Cutri} {et~al.}(2003){Cutri}, {Skrutskie}, {van Dyk}, {Beichman},
  {Carpenter}, {Chester}, {Cambresy}, {Evans}, {Fowler}, {Gizis}, {Howard},
  {Huchra}, {Jarrett}, {Kopan}, {Kirkpatrick}, {Light}, {Marsh}, {McCallon},
  {Schneider}, {Stiening}, {Sykes}, {Weinberg}, {Wheaton}, {Wheelock}, \&
  {Zacarias}}]{Cutri2003}
{Cutri}, R.~M., {Skrutskie}, M.~F., {van Dyk}, S., {et~al.} 2003, VizieR Online
  Data Catalog, II/246

\bibitem[{{Delage} {et~al.}(2022){Delage}, {Okuzumi}, {Flock}, {Pinilla}, \&
  {Dzyurkevich}}]{Delage2022}
{Delage}, T.~N., {Okuzumi}, S., {Flock}, M., {Pinilla}, P., \& {Dzyurkevich},
  N. 2022, \aap, 658, A97

\bibitem[{{Dickman}(1978)}]{Dickman1978}
{Dickman}, R.~L. 1978, \apjs, 37, 407

\bibitem[{{Dubrulle} {et~al.}(1995){Dubrulle}, {Morfill}, \&
  {Sterzik}}]{Dubrulle1995}
{Dubrulle}, B., {Morfill}, G., \& {Sterzik}, M. 1995, \icarus, 114, 237

\bibitem[{{Duffell} \& {Dong}(2015)}]{Duffell2015}
{Duffell}, P.~C. \& {Dong}, R. 2015, \apj, 802, 42

\bibitem[{{Espaillat} {et~al.}(2014){Espaillat}, {Muzerolle}, {Najita},
  {Andrews}, {Zhu}, {Calvet}, {Kraus}, {Hashimoto}, {Kraus}, \&
  {D'Alessio}}]{Espaillat2014}
{Espaillat}, C., {Muzerolle}, J., {Najita}, J., {et~al.} 2014, in Protostars
  and Planets VI, ed. H.~{Beuther}, R.~S. {Klessen}, C.~P. {Dullemond}, \&
  T.~{Henning}, 497--520

\bibitem[{{Facchini} {et~al.}(2021){Facchini}, {Teague}, {Bae}, {Benisty},
  {Keppler}, \& {Isella}}]{Facchini2021}
{Facchini}, S., {Teague}, R., {Bae}, J., {et~al.} 2021, \aj, 162, 99

\bibitem[{{Geers} {et~al.}(2006){Geers}, {Augereau}, {Pontoppidan},
  {Dullemond}, {Visser}, {Kessler-Silacci}, {Evans}, {van Dishoeck}, {Blake},
  {Boogert}, {Brown}, {Lahuis}, \& {Mer{\'\i}n}}]{Geers2006}
{Geers}, V.~C., {Augereau}, J.~C., {Pontoppidan}, K.~M., {et~al.} 2006, \aap,
  459, 545

\bibitem[{{Gregorio-Hetem} {et~al.}(1992){Gregorio-Hetem}, {Lepine}, {Quast},
  {Torres}, \& {de La Reza}}]{Gregorio-Hetem1992}
{Gregorio-Hetem}, J., {Lepine}, J.~R.~D., {Quast}, G.~R., {Torres}, C.~A.~O.,
  \& {de La Reza}, R. 1992, \aj, 103, 549

\bibitem[{{Gullbring} {et~al.}(1998){Gullbring}, {Hartmann}, {Brice{\~n}o}, \&
  {Calvet}}]{Gullbring1998}
{Gullbring}, E., {Hartmann}, L., {Brice{\~n}o}, C., \& {Calvet}, N. 1998, \apj,
  492, 323

\bibitem[{{Haffert} {et~al.}(2019){Haffert}, {Bohn}, {de Boer}, {Snellen},
  {Brinchmann}, {Girard}, {Keller}, \& {Bacon}}]{Haffert2019}
{Haffert}, S.~Y., {Bohn}, A.~J., {de Boer}, J., {et~al.} 2019, Nature
  Astronomy, 3, 749

\bibitem[{{Hashimoto} {et~al.}(2020){Hashimoto}, {Aoyama}, {Konishi}, {Uyama},
  {Takasao}, {Ikoma}, \& {Tanigawa}}]{Hashimoto2020}
{Hashimoto}, J., {Aoyama}, Y., {Konishi}, M., {et~al.} 2020, \aj, 159, 222

\bibitem[{{Hashimoto} {et~al.}(2012){Hashimoto}, {Dong}, {Kudo}, {Honda},
  {McClure}, {Zhu}, {Muto}, {Wisniewski}, {Abe}, {Brandner}, {Brandt},
  {Carson}, {Egner}, {Feldt}, {Fukagawa}, {Goto}, {Grady}, {Guyon}, {Hayano},
  {Hayashi}, {Hayashi}, {Henning}, {Hodapp}, {Ishii}, {Iye}, {Janson},
  {Kandori}, {Knapp}, {Kusakabe}, {Kuzuhara}, {Kwon}, {Matsuo}, {Mayama},
  {McElwain}, {Miyama}, {Morino}, {Moro-Martin}, {Nishimura}, {Pyo}, {Serabyn},
  {Suenaga}, {Suto}, {Suzuki}, {Takahashi}, {Takami}, {Takato}, {Terada},
  {Thalmann}, {Tomono}, {Turner}, {Watanabe}, {Yamada}, {Takami}, {Usuda}, \&
  {Tamura}}]{Hashimoto2012}
{Hashimoto}, J., {Dong}, R., {Kudo}, T., {et~al.} 2012, \apjl, 758, L19

\bibitem[{{Henkel} {et~al.}(1994){Henkel}, {Wilson}, {Langer}, {Chin}, \&
  {Mauersberger}}]{Henkel1994}
{Henkel}, C., {Wilson}, T.~L., {Langer}, N., {Chin}, Y.~N., \& {Mauersberger},
  R. 1994, in The Structure and Content of Molecular Clouds, ed. T.~L. {Wilson}
  \& K.~J. {Johnston}, Vol. 439, 72--88

\bibitem[{{Isella} {et~al.}(2019){Isella}, {Benisty}, {Teague}, {Bae},
  {Keppler}, {Facchini}, \& {P{\'e}rez}}]{Isella2019}
{Isella}, A., {Benisty}, M., {Teague}, R., {et~al.} 2019, \apjl, 879, L25

\bibitem[{{Kamp} {et~al.}(2017){Kamp}, {Thi}, {Woitke}, {Rab}, {Bouma}, \&
  {M{\'e}nard}}]{Kamp2017}
{Kamp}, I., {Thi}, W.~F., {Woitke}, P., {et~al.} 2017, \aap, 607, A41

\bibitem[{{Kamp} {et~al.}(2010){Kamp}, {Tilling}, {Woitke}, {Thi}, \&
  {Hogerheijde}}]{Kamp2010}
{Kamp}, I., {Tilling}, I., {Woitke}, P., {Thi}, W.~F., \& {Hogerheijde}, M.
  2010, \aap, 510, A18

\bibitem[{{Kanagawa} {et~al.}(2015){Kanagawa}, {Muto}, {Tanaka}, {Tanigawa},
  {Takeuchi}, {Tsukagoshi}, \& {Momose}}]{Kanagawa2015}
{Kanagawa}, K.~D., {Muto}, T., {Tanaka}, H., {et~al.} 2015, \apjl, 806, L15

\bibitem[{{Keppler} {et~al.}(2018){Keppler}, {Benisty}, {M{\"u}ller},
  {Henning}, {van Boekel}, {Cantalloube}, {Ginski}, {van Holstein}, {Maire},
  {Pohl}, {Samland }, {Avenhaus}, {Baudino}, {Boccaletti}, {de Boer},
  {Bonnefoy}, {Chauvin}, {Desidera}, {Langlois}, {Lazzoni}, {Marleau},
  {Mordasini}, {Pawellek}, {Stolker}, {Vigan}, {Zurlo}, {Birnstiel},
  {Brandner}, {Feldt}, {Flock}, {Girard}, {Gratton}, {Hagelberg}, {Isella},
  {Janson}, {Juhasz}, {Kemmer}, {Kral}, {Lagrange}, {Launhardt}, {Matter},
  {M{\'e}nard}, {Milli}, {Molli{\`e}re}, {Olofsson}, {P{\'e}rez}, {Pinilla},
  {Pinte}, {Quanz}, {Schmidt}, {Udry}, {Wahhaj}, {Williams}, {Buenzli},
  {Cudel}, {Dominik}, {Galicher}, {Kasper}, {Lannier}, {Mesa}, {Mouillet},
  {Peretti}, {Perrot}, {Salter}, {Sissa}, {Wildi}, {Abe}, {Antichi},
  {Augereau}, {Baruffolo}, {Baudoz}, {Bazzon}, {Beuzit}, {Blanchard}, {Brems},
  {Buey}, {De Caprio}, {Carbillet}, {Carle}, {Cascone}, {Cheetham}, {Claudi},
  {Costille}, {Delboulb{\'e}}, {Dohlen}, {Fantinel}, {Feautrier}, {Fusco},
  {Giro}, {Gluck}, {Gry}, {Hubin}, {Hugot}, {Jaquet}, {Le Mignant}, {Llored},
  {Madec}, {Magnard}, {Martinez}, {Maurel}, {Meyer}, {M{\"o}ller-Nilsson},
  {Moulin}, {Mugnier}, {Orign{\'e}}, {Pavlov}, {Perret}, {Petit}, {Pragt},
  {Puget}, {Rabou}, {Ramos}, {Rigal}, {Rochat}, {Roelfsema}, {Rousset}, {Roux},
  {Salasnich}, {Sauvage}, {Sevin}, {Soenke}, {Stadler}, {Suarez}, {Turatto}, \&
  {Weber}}]{Keppler2018}
{Keppler}, M., {Benisty}, M., {M{\"u}ller}, A., {et~al.} 2018, \aap, 617, A44

\bibitem[{{Keppler} {et~al.}(2019){Keppler}, {Teague}, {Bae}, {Benisty},
  {Henning}, {van Boekel}, {Chapillon}, {Pinilla}, {Williams}, {Bertrang},
  {Facchini}, {Flock}, {Ginski}, {Juhasz}, {Klahr}, {Liu}, {M{\"u}ller},
  {P{\'e}rez}, {Pohl}, {Rosotti}, {Samland}, \& {Semenov}}]{Keppler2019}
{Keppler}, M., {Teague}, R., {Bae}, J., {et~al.} 2019, \aap, 625, A118

\bibitem[{{Leemker} {et~al.}(2022){Leemker}, {Booth}, {van Dishoeck},
  {P{\'e}rez-S{\'a}nchez}, {Szul{\'a}gyi}, {Bosman}, {Bruderer}, {Facchini},
  {Hogerheijde}, {Paneque-Carre{\~n}o}, \& {Sturm}}]{Leemker2022}
{Leemker}, M., {Booth}, A.~S., {van Dishoeck}, E.~F., {et~al.} 2022, \aap, 663,
  A23

\bibitem[{{McMullin} {et~al.}(2007){McMullin}, {Waters}, {Schiebel}, {Young},
  \& {Golap}}]{McMullin2007}
{McMullin}, J.~P., {Waters}, B., {Schiebel}, D., {Young}, W., \& {Golap}, K.
  2007, in Astronomical Society of the Pacific Conference Series, Vol. 376,
  Astronomical Data Analysis Software and Systems XVI, ed. R.~A. {Shaw},
  F.~{Hill}, \& D.~J. {Bell}, 127

\bibitem[{{Mesa} {et~al.}(2019){Mesa}, {Keppler}, {Cantalloube}, {Rodet},
  {Charnay}, {Gratton}, {Langlois}, {Boccaletti}, {Bonnefoy}, {Vigan},
  {Flasseur}, {Bae}, {Benisty}, {Chauvin}, {de Boer}, {Desidera}, {Henning},
  {Lagrange}, {Meyer}, {Milli}, {M{\"u}ller}, {Pairet}, {Zurlo}, {Antoniucci},
  {Baudino}, {Brown Sevilla}, {Cascone}, {Cheetham}, {Claudi}, {Delorme},
  {D'Orazi}, {Feldt}, {Hagelberg}, {Janson}, {Kral}, {Lagadec}, {Lazzoni},
  {Ligi}, {Maire}, {Martinez}, {Menard}, {Meunier}, {Perrot}, {Petrus},
  {Pinte}, {Rickman}, {Rochat}, {Rouan}, {Samland}, {Sauvage}, {Schmidt},
  {Udry}, {Weber}, \& {Wildi}}]{Mesa2019}
{Mesa}, D., {Keppler}, M., {Cantalloube}, F., {et~al.} 2019, \aap, 632, A25

\bibitem[{{Min} {et~al.}(2009){Min}, {Dullemond}, {Dominik}, {de Koter}, \&
  {Hovenier}}]{Min2009}
{Min}, M., {Dullemond}, C.~P., {Dominik}, C., {de Koter}, A., \& {Hovenier},
  J.~W. 2009, \aap, 497, 155

\bibitem[{{M{\"u}ller} {et~al.}(2018){M{\"u}ller}, {Keppler}, {Henning},
  {Samland}, {Chauvin}, {Beust}, {Maire}, {Molaverdikhani}, {van Boekel},
  {Benisty}, {Boccaletti}, {Bonnefoy}, {Cantalloube}, {Charnay}, {Baudino},
  {Gennaro}, {Long}, {Cheetham}, {Desidera}, {Feldt}, {Fusco}, {Girard},
  {Gratton}, {Hagelberg}, {Janson}, {Lagrange}, {Langlois}, {Lazzoni}, {Ligi},
  {M{\'e}nard}, {Mesa}, {Meyer}, {Molli{\`e}re}, {Mordasini}, {Moulin},
  {Pavlov}, {Pawellek}, {Quanz}, {Ramos}, {Rouan}, {Sissa}, {Stadler}, {Vigan},
  {Wahhaj}, {Weber}, \& {Zurlo}}]{Muller2018}
{M{\"u}ller}, A., {Keppler}, M., {Henning}, T., {et~al.} 2018, \aap, 617, L2

\bibitem[{{Natta} {et~al.}(2004){Natta}, {Testi}, {Muzerolle}, {Randich},
  {Comer{\'o}n}, \& {Persi}}]{Natta2004}
{Natta}, A., {Testi}, L., {Muzerolle}, J., {et~al.} 2004, \aap, 424, 603

\bibitem[{{Paardekooper} {et~al.}(2022){Paardekooper}, {Dong}, {Duffell},
  {Fung}, {Masset}, {Ogilvie}, \& {Tanaka}}]{Paardekooper2022}
{Paardekooper}, S.-J., {Dong}, R., {Duffell}, P., {et~al.} 2022, arXiv
  e-prints, arXiv:2203.09595

\bibitem[{{Paardekooper} \& {Mellema}(2004)}]{Paardekooper2004}
{Paardekooper}, S.~J. \& {Mellema}, G. 2004, \aap, 425, L9

\bibitem[{{Pinilla} {et~al.}(2012){Pinilla}, {Birnstiel}, {Ricci}, {Dullemond},
  {Uribe}, {Testi}, \& {Natta}}]{Pinilla2012}
{Pinilla}, P., {Birnstiel}, T., {Ricci}, L., {et~al.} 2012, \aap, 538, A114

\bibitem[{{Pinte} {et~al.}(2018){Pinte}, {Price}, {M{\'e}nard}, {Duch{\^e}ne},
  {Dent}, {Hill}, {de Gregorio-Monsalvo}, {Hales}, \& {Mentiplay}}]{Pinte2018}
{Pinte}, C., {Price}, D.~J., {M{\'e}nard}, F., {et~al.} 2018, \apjl, 860, L13

\bibitem[{{Pinte} {et~al.}(2022){Pinte}, {Teague}, {Flaherty}, {Hall},
  {Facchini}, \& {Casassus}}]{Pinte2022}
{Pinte}, C., {Teague}, R., {Flaherty}, K., {et~al.} 2022, arXiv e-prints,
  arXiv:2203.09528

\bibitem[{{Pinte} {et~al.}(2019){Pinte}, {van der Plas}, {M{\'e}nard}, {Price},
  {Christiaens}, {Hill}, {Mentiplay}, {Ginski}, {Choquet}, {Boehler},
  {Duch{\^e}ne}, {Perez}, \& {Casassus}}]{Pinte2019}
{Pinte}, C., {van der Plas}, G., {M{\'e}nard}, F., {et~al.} 2019, Nature
  Astronomy, 3, 1109

\bibitem[{{Portilla-Revelo} {et~al.}(2022){Portilla-Revelo}, {Kamp}, {Rab},
  {van Dishoeck}, {Keppler}, {Min}, \& {Muro-Arena}}]{Portilla-Revelo2022}
{Portilla-Revelo}, B., {Kamp}, I., {Rab}, C., {et~al.} 2022, \aap, 658, A89

\bibitem[{{Rice} {et~al.}(2006){Rice}, {Armitage}, {Wood}, \&
  {Lodato}}]{Rice2006}
{Rice}, W.~K.~M., {Armitage}, P.~J., {Wood}, K., \& {Lodato}, G. 2006, \mnras,
  373, 1619

\bibitem[{{Rosotti}(2023)}]{Rosotti2023}
{Rosotti}, G.~P. 2023, arXiv e-prints, arXiv:2302.01433

\bibitem[{{Shakura} \& {Sunyaev}(1973)}]{Shakura1973}
{Shakura}, N.~I. \& {Sunyaev}, R.~A. 1973, \aap, 500, 33

\bibitem[{{Szul{\'a}gyi} \& {Ercolano}(2020)}]{Szulagyi2020}
{Szul{\'a}gyi}, J. \& {Ercolano}, B. 2020, \apj, 902, 126

\bibitem[{Teague(2019)}]{Teague2019}
Teague, R. 2019, The Journal of Open Source Software, 4, 1632

\bibitem[{{Teague} {et~al.}(2018){Teague}, {Bae}, {Bergin}, {Birnstiel}, \&
  {Foreman-Mackey}}]{Teague2018}
{Teague}, R., {Bae}, J., {Bergin}, E.~A., {Birnstiel}, T., \& {Foreman-Mackey},
  D. 2018, \apjl, 860, L12

\bibitem[{{Thanathibodee} {et~al.}(2020){Thanathibodee}, {Molina}, {Calvet},
  {Serna}, {Bae}, {Reynolds}, {Hern{\'a}ndez}, {Muzerolle}, \&
  {Hern{\'a}ndez}}]{Thanathibodee2020}
{Thanathibodee}, T., {Molina}, B., {Calvet}, N., {et~al.} 2020, \apj, 892, 81

\bibitem[{{Thi} {et~al.}(2011){Thi}, {Woitke}, \& {Kamp}}]{Thi2011}
{Thi}, W.~F., {Woitke}, P., \& {Kamp}, I. 2011, \mnras, 412, 711

\bibitem[{{Wang} {et~al.}(2020){Wang}, {Ginzburg}, {Ren}, {Wallack}, {Gao},
  {Mawet}, {Bond}, {Cetre}, {Wizinowich}, {De Rosa}, {Ruane}, {Liu}, {Absil},
  {Alvarez}, {Baranec}, {Choquet}, {Chun}, {Defr{\`e}re}, {Delorme},
  {Duch{\^e}ne}, {Forsberg}, {Ghez}, {Guyon}, {Hall}, {Huby}, {Jolivet},
  {Jensen-Clem}, {Jovanovic}, {Karlsson}, {Lilley}, {Matthews}, {M{\'e}nard},
  {Meshkat}, {Millar-Blanchaer}, {Ngo}, {Orban de Xivry}, {Pinte}, {Ragland},
  {Serabyn}, {Catal{\'a}n}, {Wang}, {Wetherell}, {Williams}, {Ygouf}, \&
  {Zuckerman}}]{Wang2020}
{Wang}, J.~J., {Ginzburg}, S., {Ren}, B., {et~al.} 2020, \aj, 159, 263

\bibitem[{{Wang} {et~al.}(2021){Wang}, {Vigan}, {Lacour}, {Nowak}, {Stolker},
  {De Rosa}, {Ginzburg}, {Gao}, {Abuter}, {Amorim}, {Asensio-Torres},
  {Baub{\"o}ck}, {Benisty}, {Berger}, {Beust}, {Beuzit}, {Blunt}, {Boccaletti},
  {Bohn}, {Bonnefoy}, {Bonnet}, {Brandner}, {Cantalloube}, {Caselli},
  {Charnay}, {Chauvin}, {Choquet}, {Christiaens}, {Cl{\'e}net}, {Coud{\'e} Du
  Foresto}, {Cridland}, {de Zeeuw}, {Dembet}, {Dexter}, {Drescher}, {Duvert},
  {Eckart}, {Eisenhauer}, {Facchini}, {Gao}, {Garcia}, {Garcia Lopez},
  {Gardner}, {Gendron}, {Genzel}, {Gillessen}, {Girard}, {Haubois},
  {Hei{\ss}el}, {Henning}, {Hinkley}, {Hippler}, {Horrobin}, {Houll{\'e}},
  {Hubert}, {Jim{\'e}nez-Rosales}, {Jocou}, {Kammerer}, {Keppler}, {Kervella},
  {Meyer}, {Kreidberg}, {Lagrange}, {Lapeyr{\`e}re}, {Le Bouquin}, {L{\'e}na},
  {Lutz}, {Maire}, {M{\'e}nard}, {M{\'e}rand}, {Molli{\`e}re}, {Monnier},
  {Mouillet}, {M{\"u}ller}, {Nasedkin}, {Ott}, {Otten}, {Paladini}, {Paumard},
  {Perraut}, {Perrin}, {Pfuhl}, {Pueyo}, {Rameau}, {Rodet},
  {Rodr{\'\i}guez-Coira}, {Rousset}, {Scheithauer}, {Shangguan}, {Shimizu},
  {Stadler}, {Straub}, {Straubmeier}, {Sturm}, {Tacconi}, {van Dishoeck},
  {Vincent}, {von Fellenberg}, {Ward-Duong}, {Widmann}, {Wieprecht},
  {Wiezorrek}, {Woillez}, \& {Gravity Collaboration}}]{Wang2021}
{Wang}, J.~J., {Vigan}, A., {Lacour}, S., {et~al.} 2021, \aj, 161, 148

\bibitem[{{Wilson} \& {Rood}(1994)}]{Wilson1994}
{Wilson}, T.~L. \& {Rood}, R. 1994, \araa, 32, 191

\bibitem[{{Woitke} {et~al.}(2009){Woitke}, {Kamp}, \& {Thi}}]{Woitke2009}
{Woitke}, P., {Kamp}, I., \& {Thi}, W.~F. 2009, \aap, 501, 383

\bibitem[{{Woitke} {et~al.}(2016){Woitke}, {Min}, {Pinte}, {Thi}, {Kamp},
  {Rab}, {Anthonioz}, {Antonellini}, {Baldovin-Saavedra}, {Carmona}, {Dominik},
  {Dionatos}, {Greaves}, {G{\"u}del}, {Ilee}, {Liebhart}, {M{\'e}nard},
  {Rigon}, {Waters}, {Aresu}, {Meijerink}, \& {Spaans}}]{Woitke2016}
{Woitke}, P., {Min}, M., {Pinte}, C., {et~al.} 2016, \aap, 586, A103

\bibitem[{{W{\"o}lfer} {et~al.}(2023){W{\"o}lfer}, {Facchini}, {van der Marel},
  {van Dishoeck}, {Benisty}, {Bohn}, {Francis}, {Izquierdo}, \&
  {Teague}}]{Wolfer2023}
{W{\"o}lfer}, L., {Facchini}, S., {van der Marel}, N., {et~al.} 2023, \aap,
  670, A154

\bibitem[{{Yang} {et~al.}(2012){Yang}, {Herczeg}, {Linsky}, {Brown},
  {Johns-Krull}, {Ingleby}, {Calvet}, {Bergin}, \& {Valenti}}]{Yang2012}
{Yang}, H., {Herczeg}, G.~J., {Linsky}, J.~L., {et~al.} 2012, \apj, 744, 121

\bibitem[{{Zurlo} {et~al.}(2020){Zurlo}, {Cugno}, {Montesinos}, {Perez},
  {Canovas}, {Casassus}, {Christiaens}, {Cieza}, \& {Huelamo}}]{Zurlo2020}
{Zurlo}, A., {Cugno}, G., {Montesinos}, M., {et~al.} 2020, \aap, 633, A119

\end{thebibliography}

\begin{appendix} 
\onecolumn

\section{Physical structure of the modelled protoplanetary disk around PDS 70.}
\label{sect:2D_structure}
\begin{multicols}{2}
%In this section we show the vertical cut in the three-dimensional structure of some relevant quantities of our model for the PDS 70 disk. In Fig. \ref{fig:fig} we report the cuts in the density structure and the equilibrium temperature. In the absence of a planetary companion or any other localised substructure, these fields are azimuthally symmetric. The dust opacities for each zone are shown in Fig. \ref{fig:opacities}.   
\end{multicols}

\begin{figure*}[h!]
\centering
\includegraphics[width=0.7\linewidth]{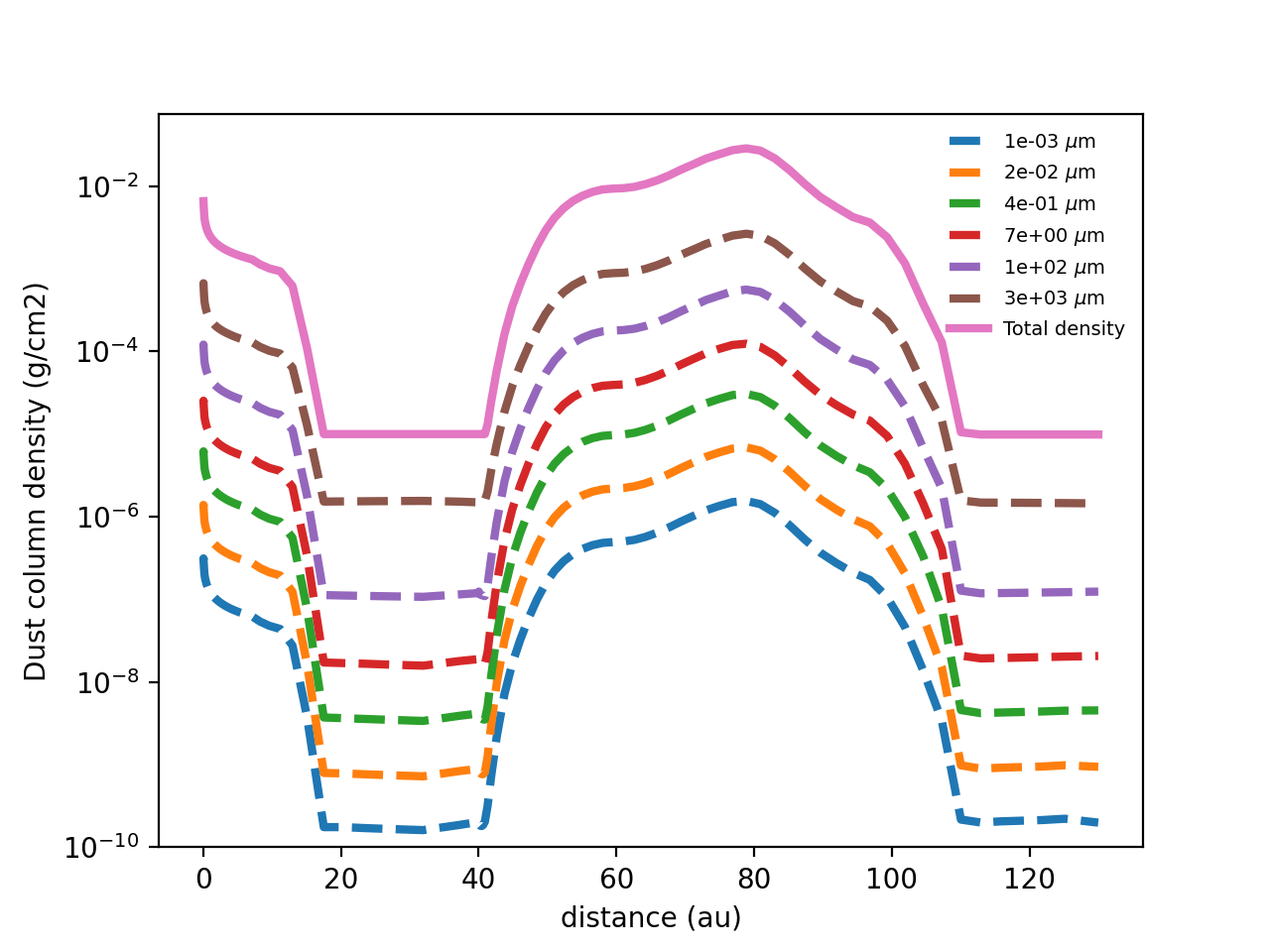}
\caption{\small Dust column density for six selected grain sizes in $\micron$ and the total column density resulting from adding up the density in each dust size bin.}
\label{fig:Sigma_dust}
\end{figure*}

\begin{figure*}[h!]
\centering
\includegraphics[width=0.7\linewidth]{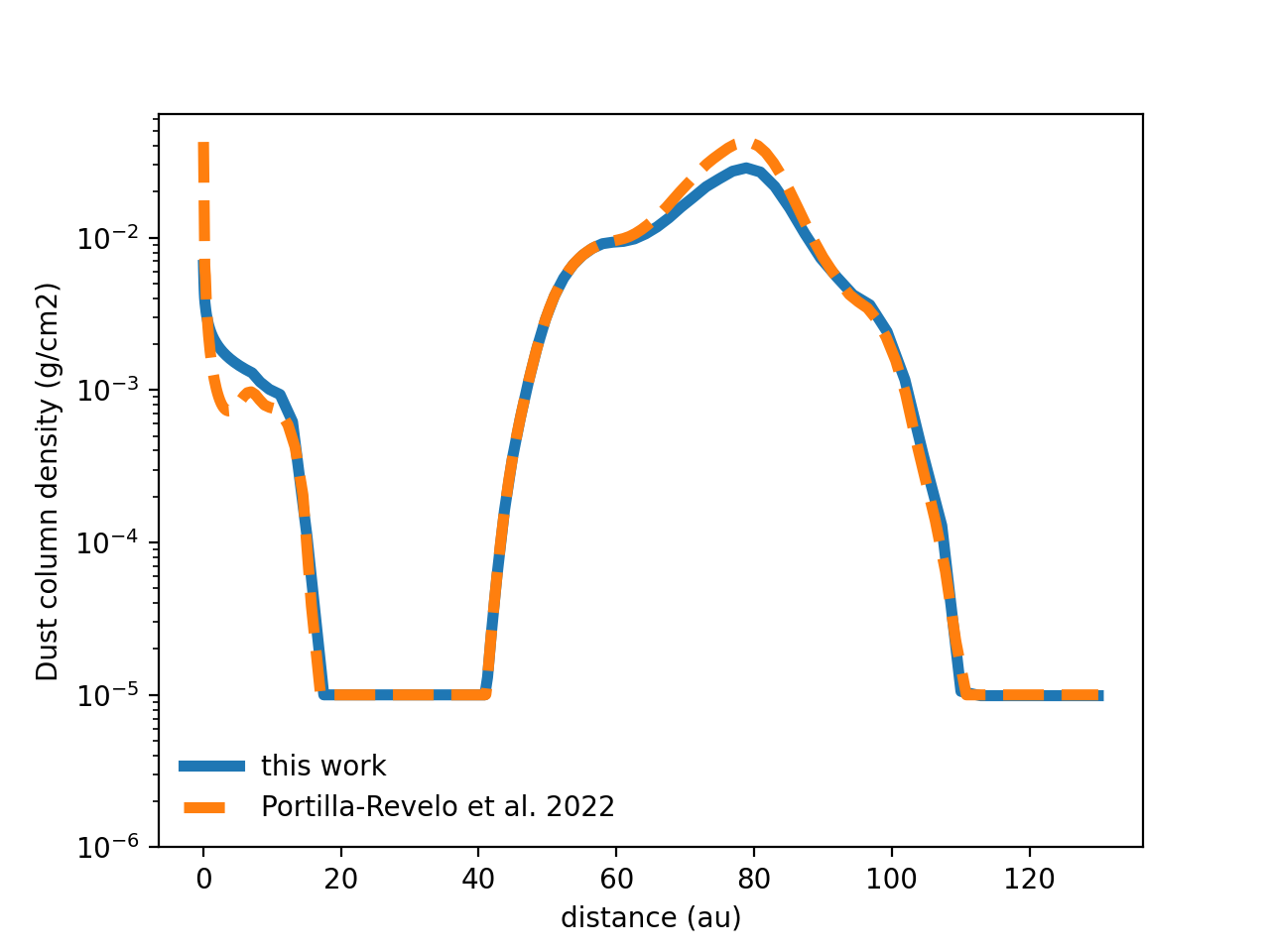}
\caption{\small Comparison of the dust column densities used in this work and that in \cite{Portilla-Revelo2022}.}
\label{fig:sigma_dust_compara_PR2022}
\end{figure*}

\begin{figure*}[h!]
\centering
\includegraphics[width=0.7\linewidth]{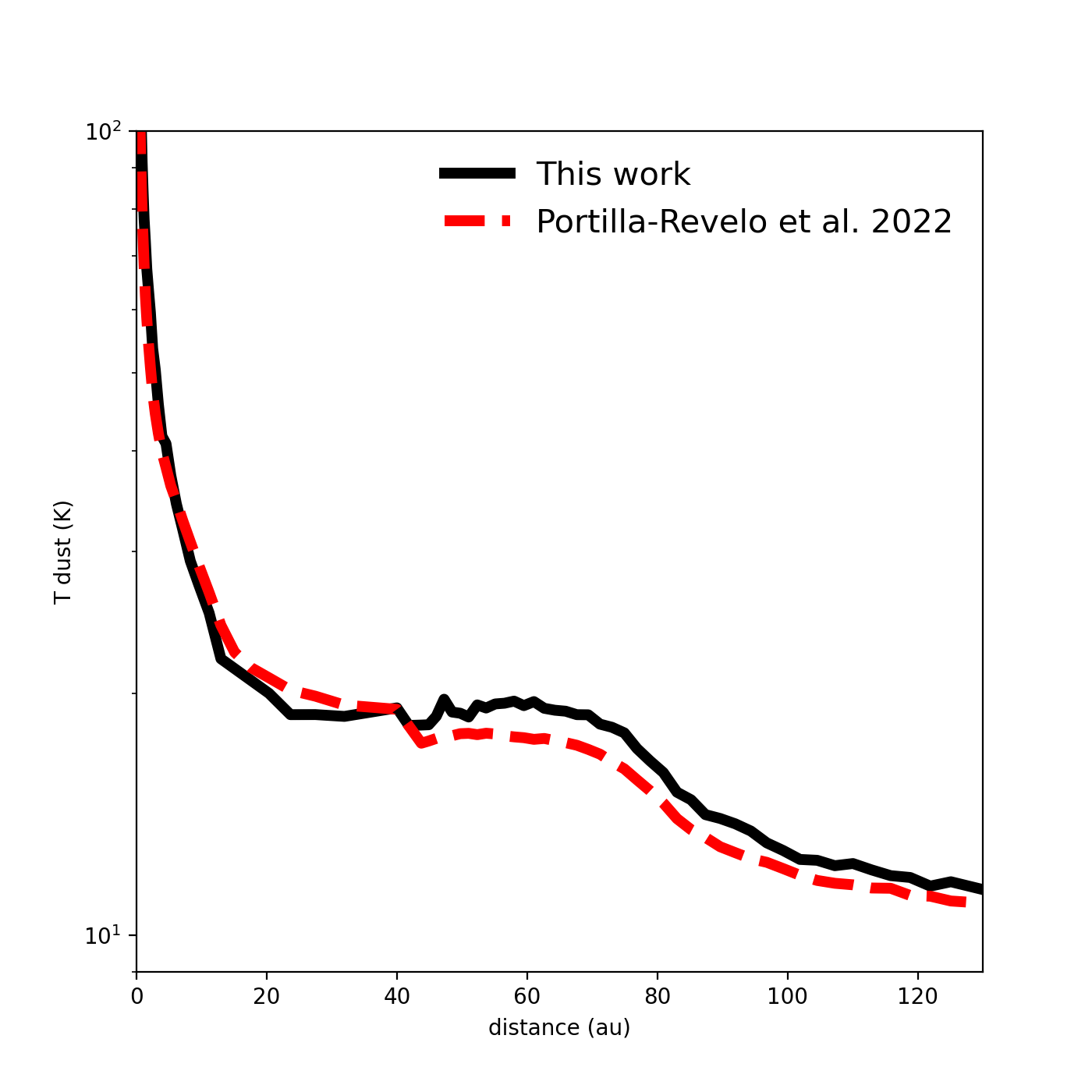}
\caption{\small Comparison of the dust midplane temperature computed in this work and that in \cite{Portilla-Revelo2022}.}
\label{fig:Td_midplane}
\end{figure*}

\begin{figure*}[h!]
\centering
\includegraphics[width=0.7\linewidth]{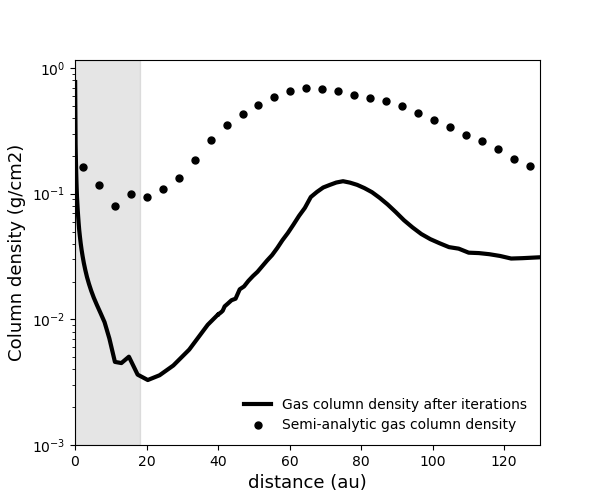}
\caption{\small Comparison of the initial gas density obtained from the semi-analytic model (Eq.\ref{eq:Tb_vs_Sigma}) and the final profile of the best representative model after the iterative procedure.}
\label{fig:Sigma_gas_compara}
\end{figure*}

\begin{figure*}[!htb]
\minipage{0.32\textwidth}
  \includegraphics[width=\linewidth]{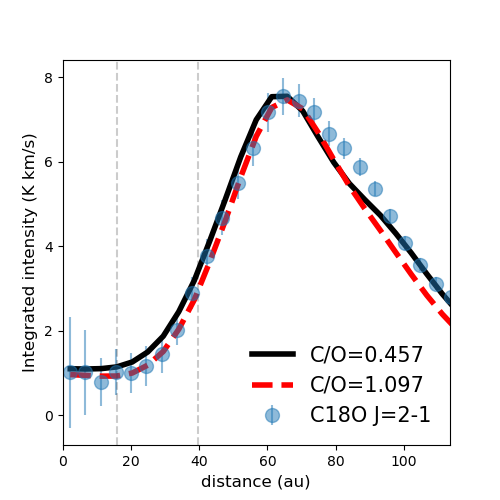}
  \label{fig:awesome_image1}
\endminipage\hfill
\minipage{0.32\textwidth}
  \includegraphics[width=\linewidth]{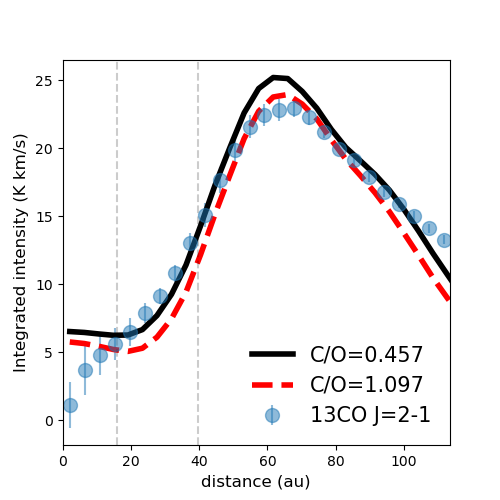}
  \label{fig:awesome_image2}
\endminipage\hfill
\minipage{0.32\textwidth}%
  \includegraphics[width=\linewidth]{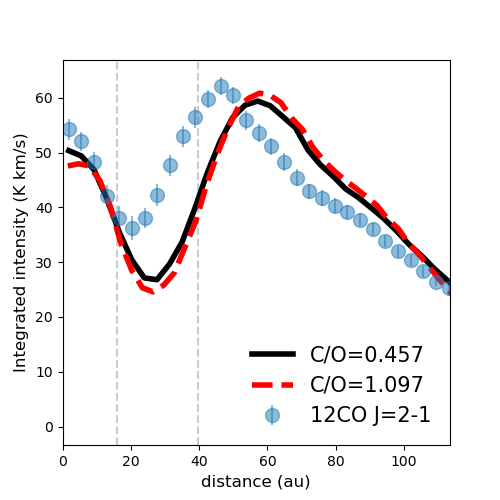}
  \label{fig:awesome_image3}
\endminipage

\caption{\small Azimuthally averaged profiles for \COO{18}{2}{1} (left), \CO{13}{2}{1} (middle) and \CO{12}{2}{1} (right) moment zero maps. The modelled curves are results from two models with different C/O ratios in the disk; the black solid lines are retrieved from the best representative model that uses $\mathrm{C/O}=0.457$ and the red dashed lines are simulations with $\mathrm{C/O}=1.097$. Vertical dashed lines indicate the limits of the dust gap.}
\label{fig:C2O_variation}
\end{figure*}

\begin{figure*}[!htb]
\minipage{\textwidth}
  \includegraphics[width=\linewidth]{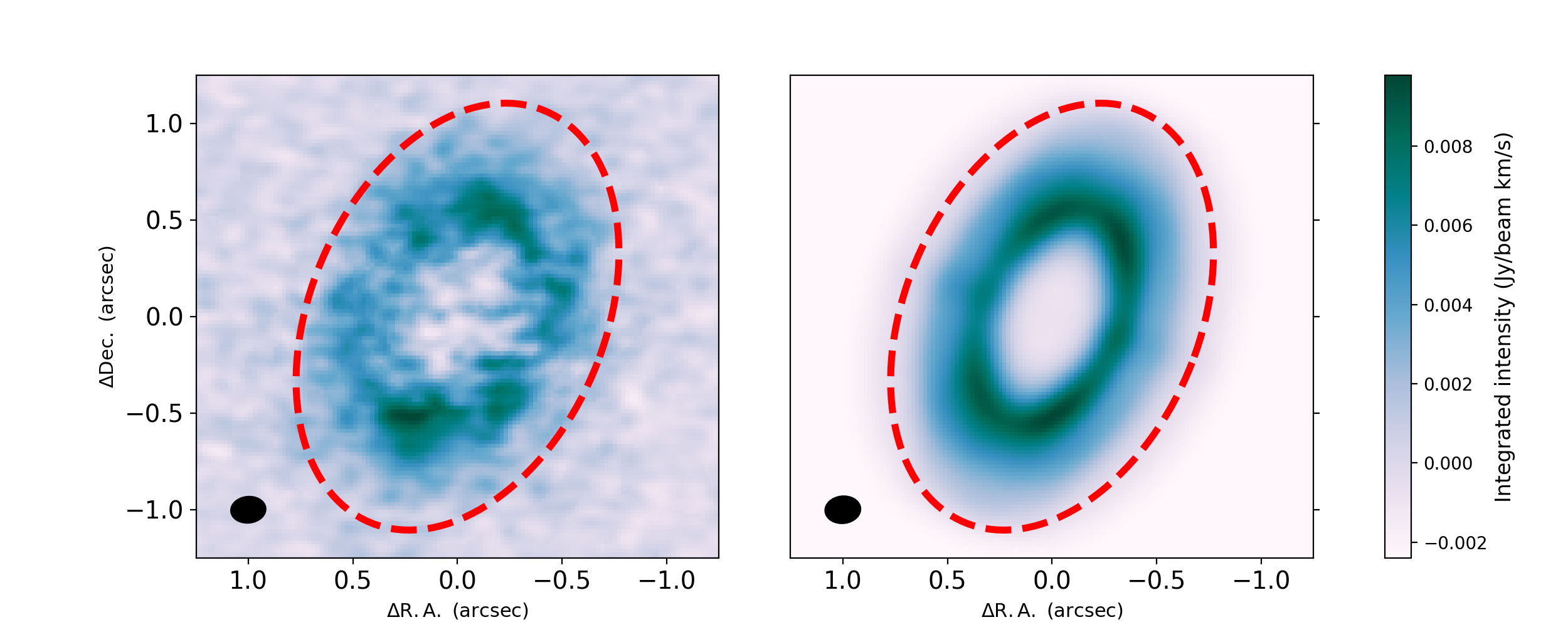}
  \label{fig:awesome_image1}
\endminipage\vfill
\minipage{\textwidth}
  \includegraphics[width=\linewidth]{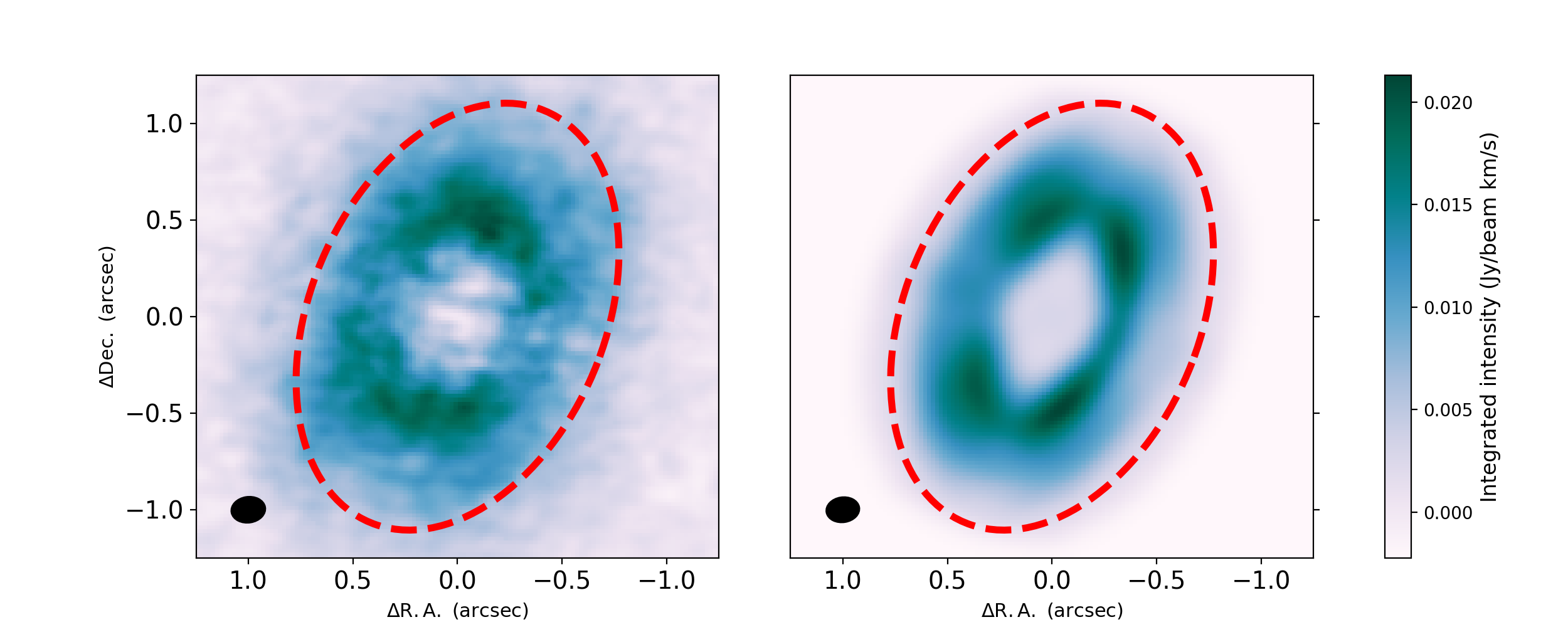}
  \label{fig:awesome_image2}
\endminipage\vfill
\minipage{\textwidth}%
  \includegraphics[width=\linewidth]{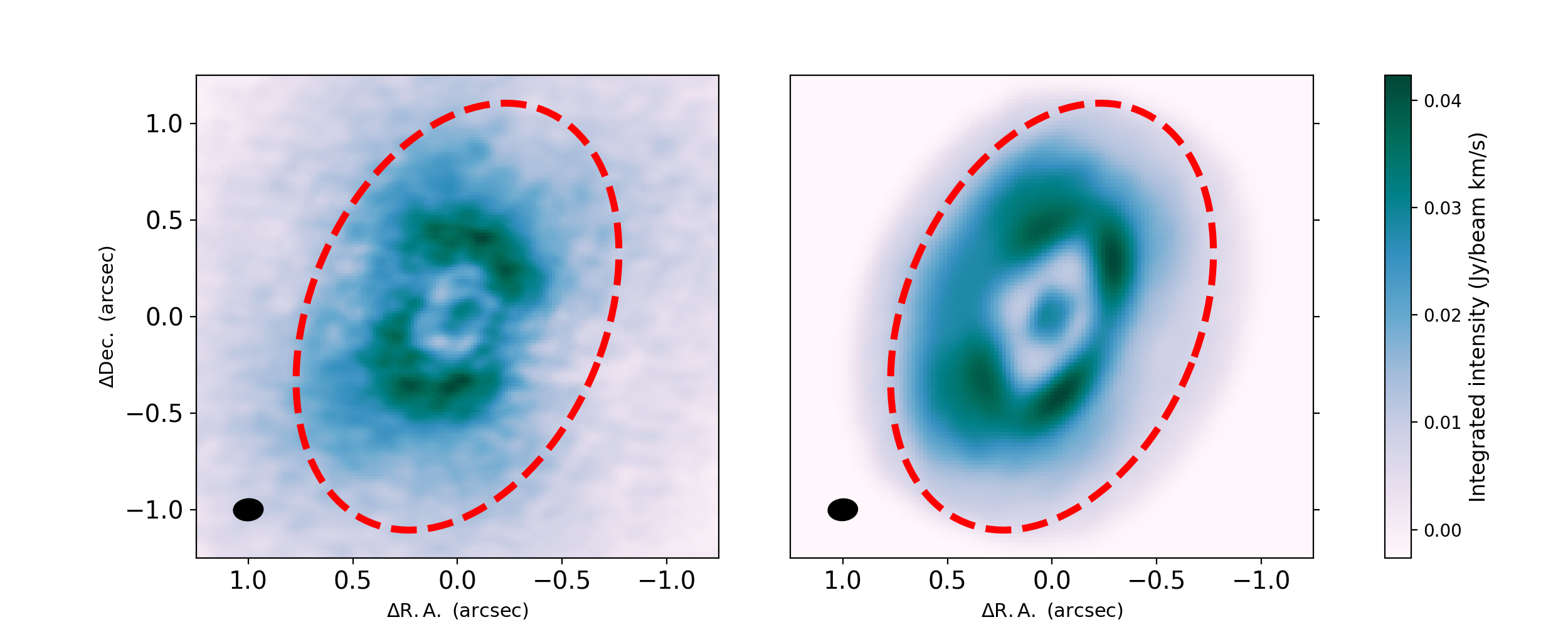}
  \label{fig:awesome_image3}
\endminipage

\caption{\small Moment zero maps for \COO{18}{2}{1} (top), \CO{13}{2}{1} (middle), and \CO{12}{2}{1} (bottom). Left column contains the observations, right column contains the models. The dashed ellipse indicates a semi-major axis of $130$ au which is the limit of applicability of our model. An orbital inclination of $51.7^\circ$ is assumed, the same as for the continuum.}
\label{fig:M0_compara}
\end{figure*}

\end{appendix}

\end{document}